\documentclass[a4paper,12pt]{article}

\usepackage[margin=1in]{geometry} 
\usepackage{setspace}

\usepackage{amsmath}
\usepackage{amsthm}
\usepackage{amssymb}
 \usepackage{siunitx} 
  \usepackage{comment} 
 \usepackage{subcaption}

\usepackage[utf8]{inputenc}
\usepackage{hyperref}
\hypersetup{
	unicode,
	pdfauthor={Author One, Author Two, Author Three},
	pdftitle={A simple article template},
	pdfsubject={A simple article template},
	pdfkeywords={article, template, simple},
	pdfproducer={LaTeX},
	pdfcreator={pdflatex}
}

\usepackage[sort&compress,numbers,square]{natbib}
\usepackage{graphicx, color}
\graphicspath{{fig/}}

\usepackage{algorithm, algpseudocode} 
\usepackage{mathrsfs} 

\usepackage{lipsum}
\usepackage{soul}

\title{ 
    On the influence of the heat transfer at the free surface of a thermally-driven rotating annulus
}
\author{Gabriel Meletti$^1$ \and Stéphane Abide$^2$ \and  Uwe Harlander $^3$ \and  Isabelle Raspo$^4$ \and Stéphane Viazzo$^4$}

\date{ 
    $^1$ Universitat Politècnica de Catalunya, \\ Departament de Matemàtiques, Barcelona, Spain \\ 
    \texttt{gabriel.meletti@upc.edu}\\
    $^2$ Université Côte d’Azur, CNRS, LJAD, France \\ \texttt{stephane.abide@univ-cotedazur.fr} \\ 
    $^3$ Brandenburg University of Technology (BTU) Cottbus-Senftenberg, Department of Aerodynamics and Fluid Mechanics, Cottbus, Germany
    \\ \texttt{uwe.harlander@b-tu.de}\\ 
    $^4$ Aix Marseille Univ, CNRS, Centrale Med, M2P2, Marseille, France \\ \texttt{\{stephane.viazzo, isabelle.raspo\}@univ-amu.fr }\\[2ex]%
     }

\begin{document}
	\maketitle

	\begin{abstract}
 
Experiments on rotating annuli that are differentially heated in the radial direction have been largely contributing to a better understanding of baroclinic instabilities. This configuration creates waves at a laboratory scale that are related to atmospheric circulations. Pioneer studies in baroclinic tanks have shown that experiments with low aspect ratios are more suitable to reproduce small-scale inertia gravity waves, but these tanks have a larger free surface, which leads to higher interactions with its surrounding environment. Considering the heat transferred through the free surface, the present work investigates its impacts on the baroclinic instability using direct numerical simulations (DNS).

\end{abstract}


\maketitle

\doublespacing

\section{\label{sec:level1} Introduction}

\begin{figure}[htbp!]
\begin{center}
  \includegraphics[width=0.4\linewidth]{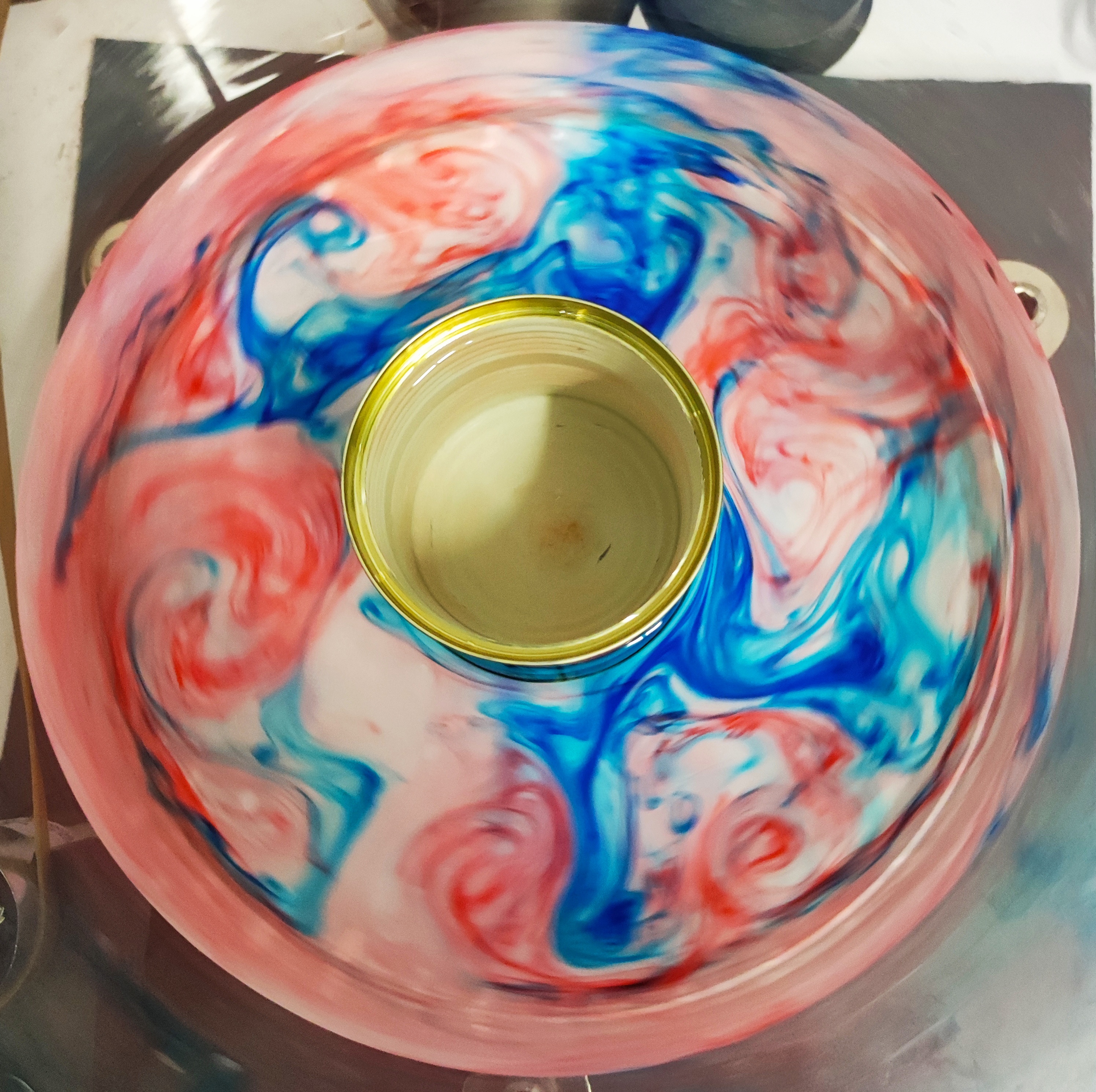}
  \end{center}
  \caption{Baroclinic instability features observed on a very simple Baroclinic Tank (BT) experiment used for lecture proposes. \label{fig:simpleBaroclinic}}
\end{figure}

Baroclinic instability is a common phenomenon in the atmosphere and oceans, driven by the presence of a temperature gradient (and a corresponding density gradient) that is not aligned with the pressure gradients. In the atmosphere and in the ocean, these instabilities can occur due to the interaction between the fluid's horizontal motion and a sloping density gradient that develops baroclinic waves, which in turn interact back with the mean flow \citep{vallis2017atmospheric}. These instabilities can lead to the formation of large-scale eddies, which play an important role in the transport of heat, momentum, nutrients, and other properties \citep{pedlosky1981resonant,read2014general}. Such eddies can be easily observed in laboratory experiments denominated Baroclinic Tanks (BTs) \citep{Harlander_Review2024}, on which the baroclinic instabilities develop when a radial temperature gradient is imposed on a fluid within a rotating annulus (that rotates in solid body rotation around the symmetry axis pointing in the axial direction). Even very simple experiments with this configuration can lead to the development of baroclinic instabilities, such as those seen in figure~\ref{fig:simpleBaroclinic}, used for lecture purpose at the Physics Laboratory of the ENS Lyon (École Normale Supérieure de Lyon). It consists of an acrylic annulus that is rotated by a small electric motor. The temperature gradient is generated by introducing ice inside a tin can that is placed at the center of the acrylic annulus, while the external wall is heated using a hair dryer. When blue ink is dropped near the inner cold surface, and red ink is dropped near the warm outer wall, we are able to see clearly the eddies of the baroclinic instability and some of its smaller scale features.   
\noindent Note that, in such BT configurations, the free surface region is not small, which facilitates heat exchange between the BT and the surrounding environment. In fact, pioneer studies such as \citet{Fultz1951experimental, hide1970baroclinic, o1969stability} have shown that Baroclinic tank experiments with low aspect ratios are more suitable to reproduce small-scale inertia gravity waves (see \citep{Harlander_Review2024} for a recent review). These tanks have a larger free surface, which leads to higher interactions with the surrounding environment. The same phenomenon can be important in ocean dynamics, for example, where the free surface plays a crucial role by allowing energy exchange between the ocean and the atmosphere. Therefore, it is relevant to explore the relationship between surface temperature and baroclinic instability, for example, for improving descriptions of geophysical phenomena such as oceanic patterns or climate and weather pattern models \citep{wunsch2004vertical}. 

\citet{scolan2017rotating} estimated the power loss at the free surface of a baroclinic tank due to evaporation, monitoring the decrease in the water level (along the depth) during experiments. The calculated power loss through evaporation was within the range of 50-80\si{\watt}. This was considered to be a significant power loss, highlighting that evaporation plays a non-negligible role in an experimental setup. These findings showed that it is important to consider heat transfer as a contributing factor in power loss during similar experiments and warrant further investigation into its effects on the overall dynamics of a rotating annulus system. Previous experimental works of \citet{bowden1965thermal} and \cite{fein1973experimental}) show different BT configurations by changing the boundary conditions at the top surface, introducing a rigid upper boundary. The upper boundaries were shown to shift the onset of the baroclinic instability, which will start at larger Burger numbers ($Bu$), defined as
\begin{equation}\label{eq:Burger}
  \textrm{Bu}=\left(\frac{N}{f}\right)^2 \left(\frac{H}{L}\right)^2,
\end{equation}
%
\noindent where $H$ is a typical vertical length scale respectively, $N$ the buoyancy frequency, $f$ the Coriolis parameter, and $L$ the gap distance defined by $L=b-a$, between the inner ($a$) and outer ($b$) radius of the BT. We note that changes in the surface temperature will lead to changes of the vertical buoyancy, $N^2=g\alpha\partial T/\partial z$ (here $g$ is the constant of gravity, $\alpha$ is the thermal expansion coefficient, $T$ the temperature, and $z$ the vertical coordinate), and consequently, to different values of $Bu$, leading to stabilizing/destabilizing effects similar to those observed in the presence of a rigid boundary.

Recently, \citet{borchert2014gravity} provided a detailed analysis of the dynamics of baroclinic instabilities, and suggested that also numerical simulations can be used completely analogous to laboratory experiments to study these surface interactions and to better understand the dynamics of the atmosphere. 
 \citet{vincze2014benchmarking} compared the signatures of baroclinic waves on a BT at its free surface with numerical simulations performed by five different groups using the same initial conditions and the same parameter regimes. They obtained similar experimental and numerical baroclinic wave patterns regarding their dominant wave modes, but showed relevant differences, for example, almost all numerical models systematically overestimated wave speeds observed experimentally. The authors also found differences in the amplitude of the temperature fluctuations when comparing experimental and numerical data. They could evaluate numerically that, not far below the free surface, the amplitudes became much larger than those evaluated at the surface. Based on previous experimental findings of \citet{rayer1998thermal}, they assumed that heat flow in the BT surface is small, i.e., that they do not have a large influence on the shape variations of baroclinic waves.
\citet{rodda2020new} also observed differences between simulations and experimental results in the surface features of a BT. In their case, they suggested that these disparities are likely attributed to variations in the upper boundary conditions, that could influence wave intensities, velocities, and wavelengths, ultimately shaping the observed features at the surface. These findings underscore the importance of accurately aligning boundary conditions between simulations and real-world experiments in order to improve the fidelity and reliability of wave studies and related geophysical phenomena. Further investigations of the impacts of heat transfer through the surface of a BT can help to better understand the specific nature of these differences between numerical and experimental investigations. This can lead not only to deeper insights into the results of experimental atmosphere-like setups but also improve the accuracy of future experimental studies in this domain.

To our knowledge, although potential impacts of heat transfer along the free surface have been mentioned several times in the literature, no systematic investigation has yet been performed.
In this study, we will therefore investigate the sensitivity of the baroclinic instability to the heat transfer at the free surface.
In other words, we will not consider an adiabatic condition at the free surface, as usual. On the contrary, we will impose different heat fluxes at the surface, and observe how they change the baroclinic instability features. 
The use of a DNS code allows us to control the boundary conditions, and obtain a quantitative analysis of the results obtained.

The paper is structured as follows: in Section~\ref{sec:SetUp}, we describe the BT geometry under consideration, along with the equations used to model the problem and the key non-dimensional parameters. In Section~\ref{subsec:DNS}, we present the DNS code and the numerical methods we employ. In Section~\ref{subsec:Validation}, we validate the code by comparing the results obtained for baroclinic cavities with other numerical methods and experimental data. Section~\ref{sec:results} presents the effects on the baroclinic instability of changing the heat flux in the BT surface. Section \ref{sec:small-scales} focuses on small-scale features, and, lastly, in Section~\ref{sec:conclusions}, we present the conclusions of this study.

\section{\label{sec:SetUp}  Flow Set-up and Governing Equations}
The present study is based on direct numerical simulations. 
In this section, we will first present the governing equations, followed by the numerical method used to solve them. We will also validate our numerical simulations by comparing results with those previously obtained both experimentally and numerically by \citet{vincze2014benchmarking}.

The BT configuration is similar to those previously considered in \cite{borchert2014gravity,rodda2018baroclinic}, on which an annulus of inner and outer radius $a<b$ is filled with water up to a height $d$, defining the dimensionless parameters $\eta=a/b$, and the vertical aspect ratio $\Gamma=H/L$.
The BT is subjected to a solid-body rotation of rate $\Omega \mathbf{e}_z$ such that the free surface remains horizontal.
The difference $\Delta T$ between the hot inner and colder outer cylinder's temperatures, respectively $T_h$ and $T_c$, is maintained sufficiently small so that the Boussinesq approximation remains valid. In the Boussinesq approximation, we consider density to be constant, except in the buoyancy and centrifugal acceleration terms in equations (\ref{eqn_div})-(\ref{eqn_ener}).
In these terms, the flow density varies linearly with temperature, considering a reference temperature $T_r=(T_h+T_c)/2$.

We use as reference length, time, and velocity, the cylinder gap $L=b-a$, the viscous diffusion time $L^2/\nu$, and the viscous velocity $\nu/L$  respectively.
Here $\nu$ and $\kappa$ are the kinematic viscosity and the thermal diffusivity (see their values on Table~\ref{tab:Params_BBT}).
This scaling leads to the following dimensionless governing equations:
\begin{equation}\label{eqn_div}
    \nabla \cdot \mathbf{u} = 0,
\end{equation}
\begin{equation}
    \partial_t \mathbf{u} 
    + \left( \mathbf{u} \cdot \nabla  \right) \mathbf{u} 
    + \sqrt{Ta} \, \mathbf{e}_z \times \mathbf{u} = 
    - \nabla  p 
    + \nabla^2 \mathbf{u}  
    + \dfrac{Ra}{Pr} \theta \mathbf{e}_z
    + Ra^\prime \theta r \mathbf{e}_r
\end{equation}
\begin{equation}\label{eqn_ener}
    \partial_t \theta 
    + \mathbf{u} \cdot \nabla  \theta
    = 
     \dfrac{1}{Pr} \nabla^2 \theta,  
\end{equation}
where $\mathbf{u}=u_r \mathbf{e}_r+u_\theta \mathbf{e}_\theta+u_z \mathbf{e}_z$ is the velocity expressed in the cylindrical coordinate system, with $(\mathbf{e}_r,\mathbf{e}_\theta,\mathbf{e}_z)$ the unit vectors in radial, azimuthal and axial directions respectively, and $\theta  = \frac{T - T_r}{T_h - T_c}$ is the dimensionless temperature deviation from the reference temperature.

The dimensionless numbers in Equations~(\ref{eqn_div}-\ref{eqn_ener}) are the Taylor ($Ta$), the Rayleigh ($Ra$ and $Ra^\prime$) and the Prandtl ($Pr$) numbers, defined as:
\begin{equation}\label{eq:non-dimension_parameters}
  Ta=\frac{4\Omega^2 (b-a)^4}{\nu^2},\quad
  Ra= \frac{ g \alpha \Delta T (b-a)^3}{\nu \kappa} ,\quad
  Ra^\prime = \frac{ \alpha \Delta T}{4}Ta ,\quad
  Pr= \frac{ \nu }{ \kappa } ,
\end{equation}
where the Taylor and Rayleigh numbers are respectively related to Coriolis and Buoyancy forces.
The second definition of the Rayleigh number ($Ra^\prime$) presented in Eq.(\ref{eq:non-dimension_parameters}) is related to the centrifugal acceleration. Taking into account the aspect ratio of the BT ($\Gamma$), we obtain a usual definition of the thermal Rossby number given by:
\begin{equation}\label{eq:RossbyThermal}
Ro_T = \frac{4 Ra }{ Pr Ta} \frac{1}{ \Gamma}.    
\end{equation}
\noindent Note that, in Equation~(\ref{eq:non-dimension_parameters}), we considered here the non-dimensionalization as in \citet{von2013influence}.
The Taylor and the Rayleigh numbers could also have been defined taking into account the BT aspect ratio ($\Gamma$) as $\widetilde{Ta}=Ta/\Gamma$, and $\widetilde{Ra}=\frac{ g \alpha \Delta T (b-a)^3}{\nu \kappa} \Gamma$. 
In our case, though, we used in our equations the definitions presented in~Eq.(\ref{eq:non-dimension_parameters}).

The influence of heat exchanges between the free surface of the tank and ambient air on the baroclinic instability dynamics is investigated here using a Direct Numerical Simulations (DNS) code, i.e., we solve the set of non-linear equations (\ref{eqn_div})-(\ref{eqn_ener}) without using turbulence models to describe the small-scale dynamics. Instead, the code has sufficiently fine spatial resolution to resolve the dissipative scales. Although using DNS codes avoids modeling the viscous dissipation in the small scales (as in Large Eddy Simulation (LES) codes), a well-resolved simulation of the baroclinic waves remains a computational challenge due to the occurrence of multiple scales that interact. This is especially true since our investigations model the same configuration presented in the experimental work of \citet{rodda2020new}. The difficulties in numerically modeling this particular BT configuration come from the fact that it has not only a large gap $L$ but also small radial temperature differences $\Delta T$. The advantage of using a numerical approach at this stage is that, differently from experiments, it will allow us to accurately control the boundary conditions at the free surface.

Regarding the boundary conditions, no-slip $\mathbf{u} \cdot \mathbf{n}=0$ is considered at all walls except the upper surface of the BT. At this free surface, stress-free boundary conditions are considered, i.e., $\partial_{\mathbf{n}}u_r=\partial_{\mathbf{n}}u_\phi=0$ and $u_z=0$.
At the inner and outer cylinders, the dimensionless temperature is set to $\Theta_i=-1/2$ and $\Theta_o=+1/2$, respectively. The bottom wall of the tank is considered to be adiabatic, i.e., $\partial_z\Theta=0$. In contrast, at the free surface, we consider heat transfer with the surrounding environment given by Newton's law on a dimensionless formulation, reading:
\begin{equation}\label{eq:heat_transf}
    -\partial_{z} \Theta = Bi (\Theta-\Theta_q), 
\end{equation}
where $Bi=hL/\lambda$ is the Biot number, and $\Theta_q$ is the dimensionless temperature of the ambient air that is in contact with the free water surface, here named quiescent temperature. The Biot number is based on the thermal conductivity of the flow ($\lambda$) and the effective heat transfer coefficient ($h$), set to $h=5 WK^ {-1} m^ {-2} $, which is a typical value in natural convection.
Considering the boundary conditions described, the baroclinic cavity is completely described by the set of dimensionless parameters $(Ta, Ra, Ra^\prime, Bi, \Theta_q, Pr, \eta, \Gamma)$. The ranges of these dimensionless parameters that were considered in this work are summarized in Table~\ref{tab:Params_BBT}.
\begin{table*}
  \centering
  \begin{tabular}{lcc}
    \hline
   Parameters  & values & Non-dimensional associated \\
    \hline
    Inner radius, $[m]$ & $a=0.35$ &  \multicolumn{1}{c}{$\eta=0.5$}\\
    Outer radius, $[m]$ & $b=0.70$ &  \multicolumn{1}{c}{$\Gamma\simeq0.1714$} \\
    Fluid depth, $[m]$ & $d=0.04$ &    \\
    \hline
    kinematic viscosity , $[m^2s^{-1}]$ & $\nu=1.004\times10^{-6}$ &  \multicolumn{1}{c}{$\mathrm{Pr}=7$}  \\
    thermal diffusivity, $[m^2s^{-1}]$ & $\kappa =1.4340 \times 10^{-7}$ &  \\
    thermal conductivity, $[ms^{-1}]$ & $\lambda =0.6$ &   \\
    thermal expansion , $[K^{-1}]$ & $\alpha =2.07 \times 10^{-4}$ &   \\
    \hline
    \vspace{0.2cm}
    Angular velocity, $[rpm]$ & $0.3 \leq \Omega \leq 0.6 $
                                      &  \begin{tabular}{@{}c@{}} $ \{ 5.88 \times 10^{7} \leq \mathrm{Ta} \leq  2.35 \times 10^{8}$; \\ $3.43 \times 10^{8} \leq \widetilde{\mathrm{Ta}} \leq  1.37 \times 10^{9} \} $ \end{tabular} \\ \vspace{0.2cm}
                                      %
                                      %
    Inner wall temperature, $[^\circ C]$ & $T_c =20.0$ &  \begin{tabular}{@{}c@{}} $ \{ Ra=1.51 \times 10^{9} $; \\ $ 0.63 \leq \mathrm{Ro}_T \leq 2.52 \}$ \end{tabular} \\  \vspace{0.2cm}
    Outer wall temperature, $[^\circ C]$ & $T_h= 22.5$ & $ 7.60\time 10^{3} \leq \mathrm{Ra}^\prime \leq 3.04\time 10^{4}$  \\
    Quiescent temperature, $[^\circ C]$ & $ 13.75 \leq T_q \leq 28.75 $ & $ -3 \leq \Theta_q \leq +3$ \\
    heat transfer coefficient $h$  , $[WK^{-1}]$ & $5$ & $ \mathrm{Bi}=2.91$ \\
    \hline %
  \end{tabular}
  \caption{Physical and dimensionless parameters for an atmosphere-like BT used in {\cite{rodda2020new}}.}
  \label{tab:Params_BBT}
\end{table*}
%



\subsection{\label{subsec:DNS} Numerical methods}

As previously mentioned, DNS has been retained to investigate the impacts of the surface heat flux on the baroclinic wave features. The Navier-Stokes equations coupled with the energy equation in (\ref{eqn_div})-(\ref{eqn_ener}) are then solved using a higher-order numerical method combined with parallel computation.
The time advancement of the equations is performed using a semi-explicit Runge-Kutta 3/Crank-Nicolson scheme, and a projection algorithm is used to handle the resulting pressure-velocity coupling.
Spatial discretization is done with a fourth-order compact scheme in the radial and axial directions, and a Fourier-Galerkin scheme in the azimuthal direction.
The linear systems resulting from this discretization are solved by a direct solver known as successive diagonalizations.
This approach combines high-performance computing and higher-order discretization on staggered grids for cylindrical coordinate systems.
The numerical method is detailed and validated in \citet{abide2017,abide2018} and it was already applied in other investigations of stratified rotating annular flows in \citet{meletti2021experiments,meletti2023parameter}.

\section{\label{subsec:Validation}Code validation for baroclinic cavity flows}
%
In this section, we will present a specific validation of our DNS solver to describe the characteristics of baroclinic waves. We consider BT configurations  previously used in the experimental studies of \citet{vincze2014benchmarking,rodda2020new}, and confront the results with the numerical benchmark that was also presented in \cite{vincze2014benchmarking}.

To begin, we conducted simulations for the various test cases of the baroclinic configuration presented in the left column of Table~\ref{tab:Param_Values}, called the small tank experimental configuration. The first cavity geometry we considered had an aspect ratio of $\Gamma=1.8$. For obtaining an unstable baroclinic regime with a standing wave pattern, we selected a radial temperature difference of $\Delta T=\SI{8}{\kelvin}$, and rotation rates $\Omega=7$, $9$, $11$ and $17\,$ \text{rpm}.
\begin{figure}[htbp!]
  \includegraphics[width=0.8\linewidth]{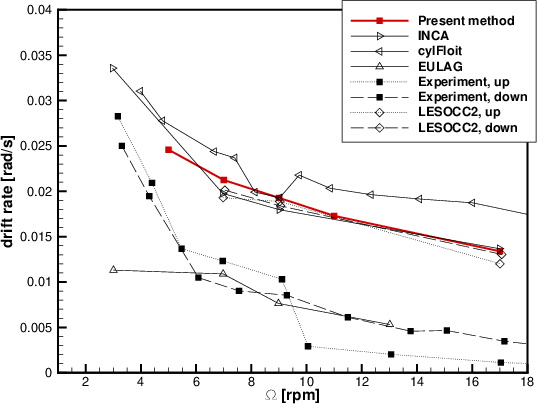}
  \caption{Drift rate of the dominant wave mode as a function of rotation rate $\Omega$.\label{fig:drift_vs_omega_small_tank}}
\end{figure}

To evaluate the outcomes of our simulation, we compare the drift rate and thermal variability with experimental data and previous numerical simulations presented in \citet{vincze2014benchmarking}. 
To obtain these parameters, we performed a harmonic analysis of the free surface temperature at the mid-radius position, as outlined in Appendix~\ref{Appendix:HarmonicAnalysis}. Figure~\ref{fig:drift_vs_omega_small_tank} presents the correlation between drift rate and rotation rate, demonstrating that our numerical results fall within the spread range of values reported in \citet{vincze2014benchmarking}. 
\begin{figure}[htbp!]
  \includegraphics[width=1.0\linewidth]{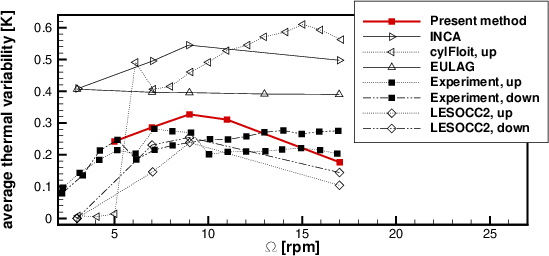}
  \caption{Average thermal variability as a function of rotation rate $\Omega$}\label{fig:sigma_vs_omega} 
\end{figure}
Notably, we observe a strong agreement with the results obtained using the codes INCA~\citep{hickel2006adaptive} and LESOCC2~\citep{hinterberger2007}. However, our simulations reveal a discrepancy when compared to experimental data and results from other solvers, underscoring the challenging nature of this benchmark problem. This observation also applies to the thermal variability $\sigma$ (defined in the appendix~\ref{Appendix:HarmonicAnalysis}), shown in Fig.~\ref{fig:sigma_vs_omega}. Despite these complexities, it is worth highlighting that our findings demonstrate good agreement with the results obtained by the LESOCC2 solver \citep{hinterberger2007}. The thermal variability resulting from our solver also agrees reasonably with experimental data. Note also that the EULAG code agrees with experimental data on the drift rate, but shows larger discrepancies with respect to the thermal variability than we observed with our code. 
Considering the spread of data for this benchmark, we can affirm that our method captures the fundamental dynamics of baroclinic waves, akin to the other DNS codes. The fact that different codes agree, but all show some discrepancies when compared to the experimental data, indicates the presence of unaccounted physical phenomena that were not captured by the models and still remain to be better comprehended.
\begin{table}
    \centering
    \begin{tabular}{|l|c|c|}
    \hline
             &  Small tank  &   Big tank  \\ \hline 
     $\eta$    & $0.37$   &  $0.50$  \\ \hline
     $\Gamma$     & $0.13-1.73$   &  $0.17$    \\ \hline
     $Ta$        & $3.44\times 10^5 - 1.38\times 10^8$      &   $6.53\times 10^6 - 9.40\times 10^8$      \\ \hline
     $Ra$        &  $4.77\times 10^7$     &    $3.03\times 10^9$     \\ \hline
     $Pr$        &  $7$     &    $7$     \\ \hline
    \end{tabular}
    \caption{Parameter values for the two baroclinic cavities considered by~\citet{rodda2020new} }
    \label{tab:Param_Values}
  \end{table}

\begin{figure*}[htbp!]
\begin{center}
  \includegraphics[width=0.3\linewidth]{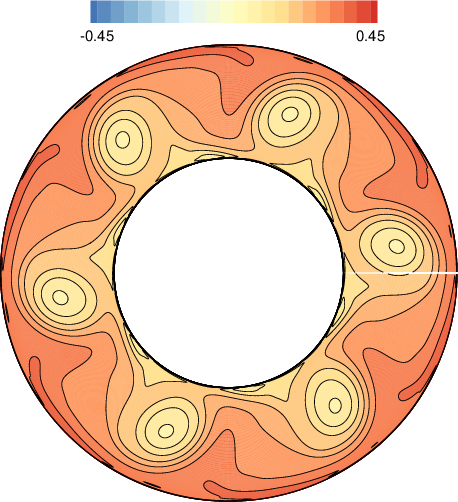}
  \includegraphics[width=0.4\linewidth]{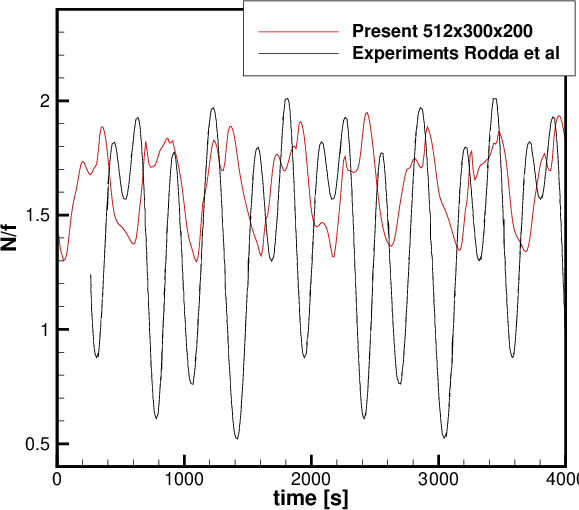}
  \end{center}
  \caption{Free surface temperature obtained by numerical simulation for $\Delta T=3.5\,\text{K}$ and $\Omega =0.7\,\text{rpm}$ (left) and the corresponding time evolution of $N/f$ (right) of the present numerical simulations confronted with the experimental results obtained by \citet{rodda2020new} .}
  \label{fig:bigtank_rodda}
\end{figure*}

For further validation, an additional simulation was performed in a baroclinic cavity with aspect ratio $\Gamma<1$.
More precisely, we consider the atmospheric-like baroclinic cavity used in the experiments of \cite{rodda2020new}, with dimensions given in the right column of Table \ref{tab:Param_Values}.
In this configuration, the notably large free surface facilitates significant heat exchanges between the fluid within the tank and the surrounding air.
The rotation rate and the radial temperature difference are set to $\Omega=0.7\,\text{rpm}$ and $\Delta T=3.5\si{\kelvin}$, for which a vacillating regime was recently observed by \citet{rodda2020new}. Figure~\ref{fig:bigtank_rodda} shows the surface temperature obtained in our numerical simulations exhibiting a final $m=6$ dominant wave number of the baroclinic instability.

In \citet{rodda2020new}, a dominant wave mode with wavenumber $m=7$ was observed experimentally. In fact, the dominant wave mode initially appeared with the same wavenumber $m=7$ in numerical simulations (not shown), before transitioning to the final $m=6$ case presented on the left side of Fig.~\ref{fig:bigtank_rodda}. On the right side of the figure, the time series of the ratio $N/f$ (buoyancy frequency $N$ to Coriolis parameter $f$) for both experimental and numerical data over a $\SI{4000}{\second}$ interval is presented.

Despite some differences in the final wavenumber between numerical and experimental results, both datasets exhibit qualitatively similar behavior. Notably, the experimental fluctuations have a larger amplitude compared to the numerical simulations, a trend also observed by \citet{rodda2020new}. This discrepancy in amplitudes and wavenumber might suggest that simulations may not fully capture the multi-scale nature of the problem. Furthermore, the somewhat irregular pattern, which is asymmetric with respect to its average value, observed in the experimental data from \citet{rodda2020new} hints at the influence of free surface effects.

To address the amplitude discrepancies between the various numerical methods and the experimental results, we employ here a fine mesh resolution of  $512 \times 300 \times 200$, a computationally expensive option, often avoided in simulations. This refined grid yields dimensionless drift rate ($c_n$) and thermal variability ($\sigma$) of $(c_n,\sigma)=(241,0.577)$, enabling a more accurate resolution of smaller scales that may account for the discrepancies in drift and thermal variability. 
Additionally, we will investigate the influence of temperature changes in the free-surface temperature on the development of the azimuthal wavenumber, a factor commonly neglected in numerical simulations. 
Note that experimental BTs inevitably experience heat exchange between their free surfaces and the surrounding environment. 
Moreover, we will examine the impact of temperature fluctuations at the free surface on the dominant wavenumber of baroclinic instability. This aspect is often neglected in numerical simulations too; however, it is relevant since real experimental BTs naturally undergo temperature exchanges with their surroundings, and these exchanges have non-negligible effects on the final obtained wave modes.

\section{Results and discussions}\label{sec:results}

\begin{figure*}[htbp!]
    \includegraphics[width=0.13\linewidth]{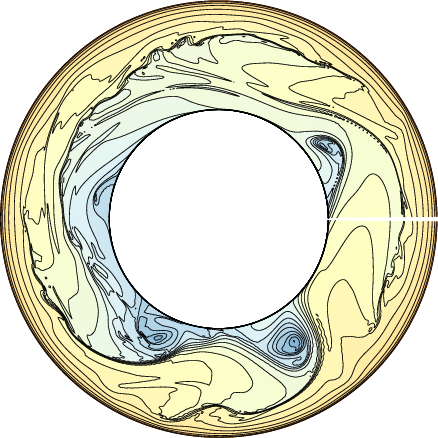}
    \includegraphics[width=0.13\linewidth]{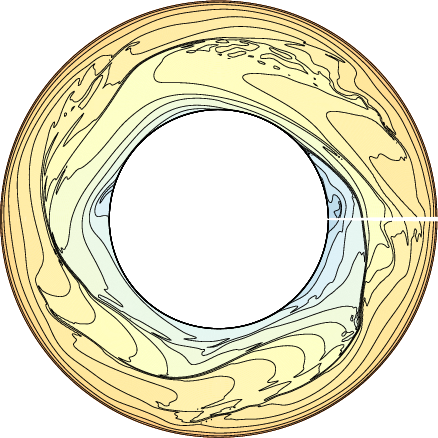}
    \includegraphics[width=0.13\linewidth]{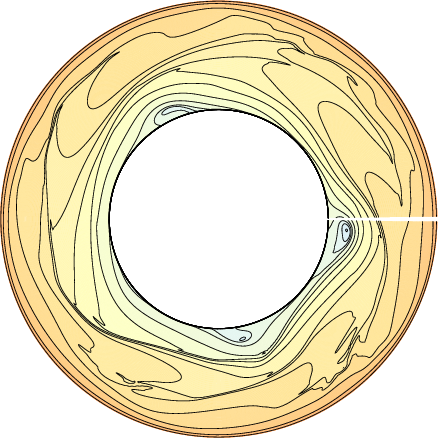}
    \includegraphics[width=0.13\linewidth]{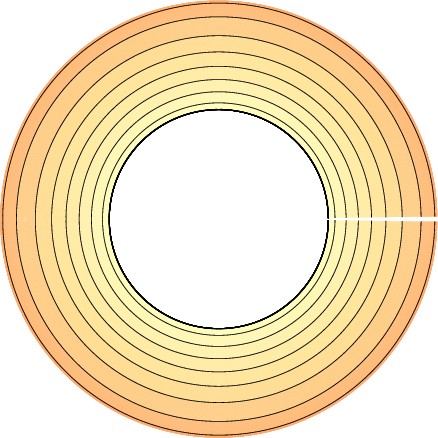}
    \includegraphics[width=0.13\linewidth]{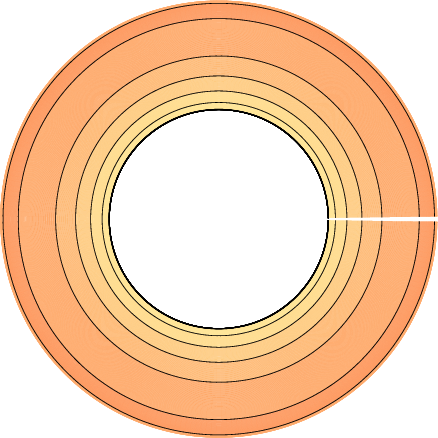}
    \includegraphics[width=0.13\linewidth]{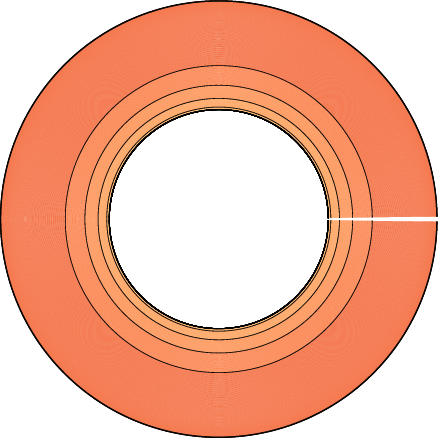}
    \includegraphics[width=0.13\linewidth]{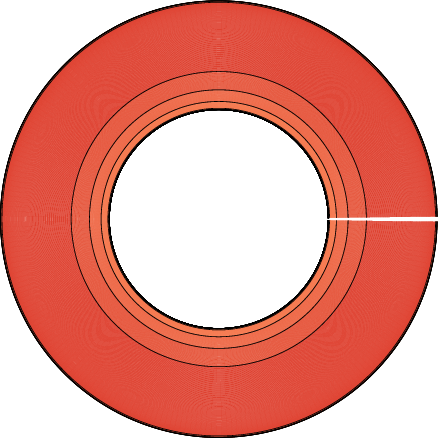}\\
    \includegraphics[width=0.13\linewidth]{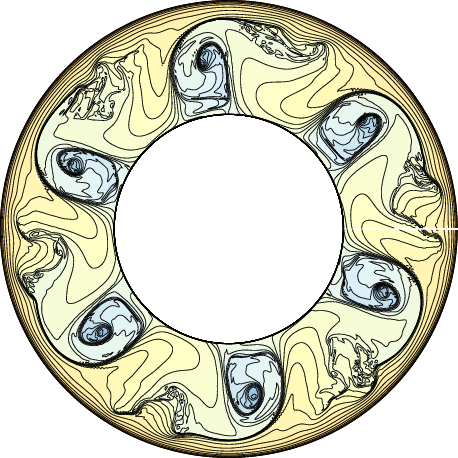}
    \includegraphics[width=0.13\linewidth]{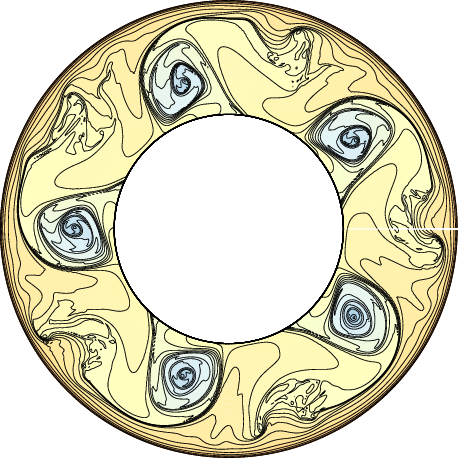}
    \includegraphics[width=0.13\linewidth]{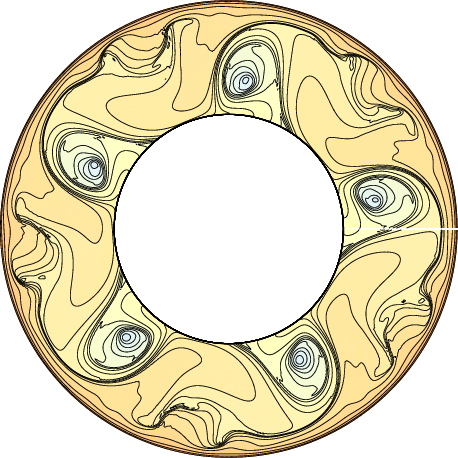}
    \includegraphics[width=0.13\linewidth]{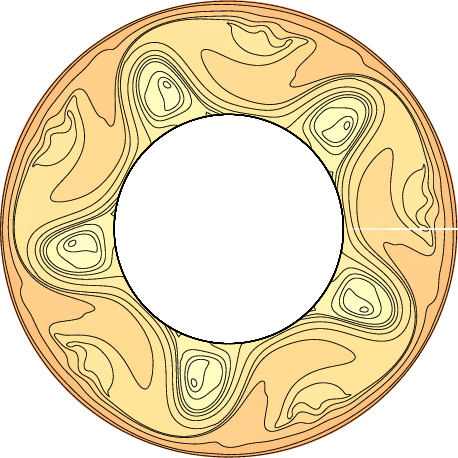}
    \includegraphics[width=0.13\linewidth]{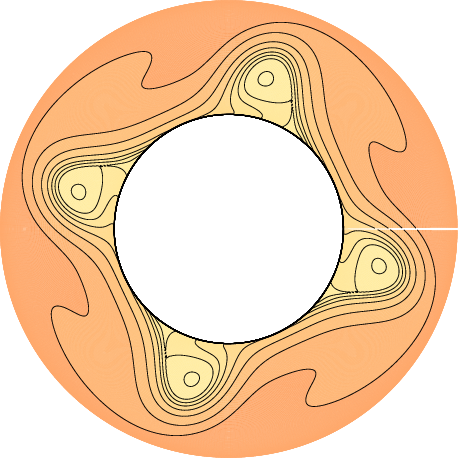}
    \includegraphics[width=0.13\linewidth]{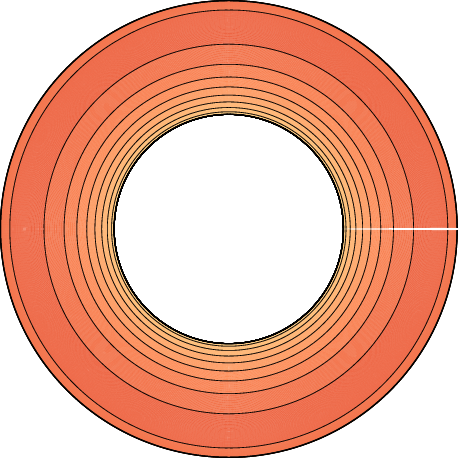}
    \includegraphics[width=0.13\linewidth]{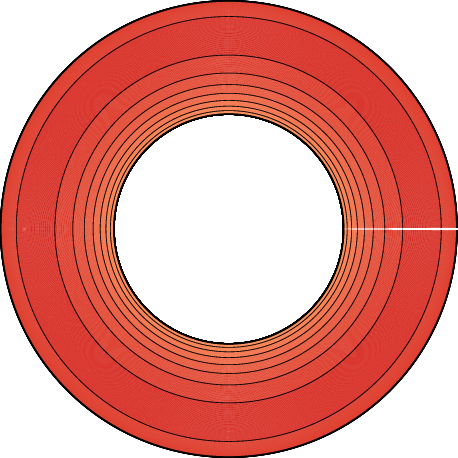}\\
    \includegraphics[width=0.13\linewidth]{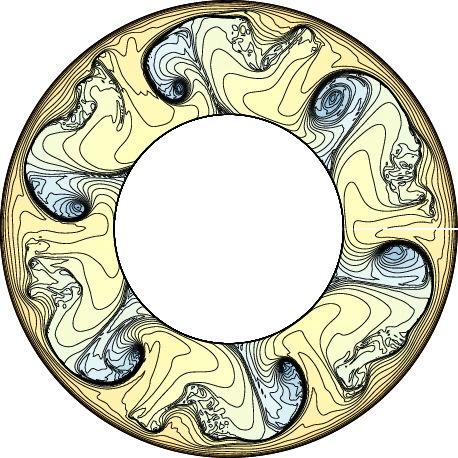}
    \includegraphics[width=0.13\linewidth]{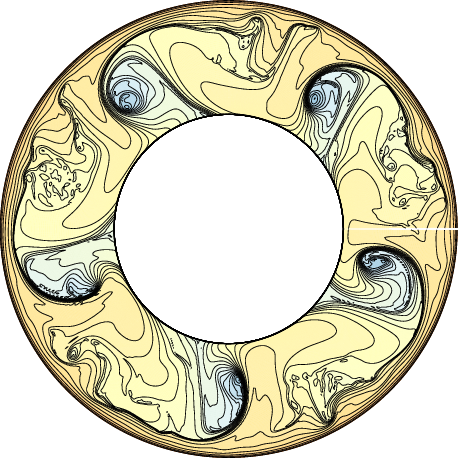}
    \includegraphics[width=0.13\linewidth]{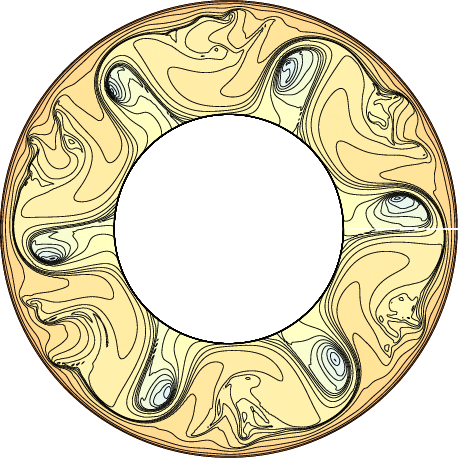}
    \includegraphics[width=0.13\linewidth]{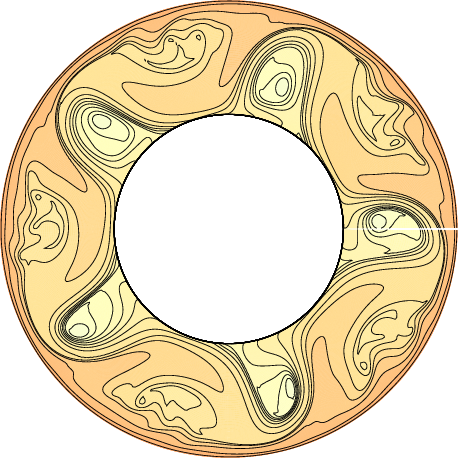}
    \includegraphics[width=0.13\linewidth]{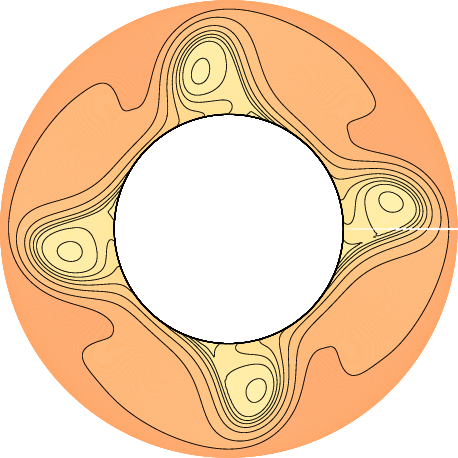}
    \includegraphics[width=0.13\linewidth]{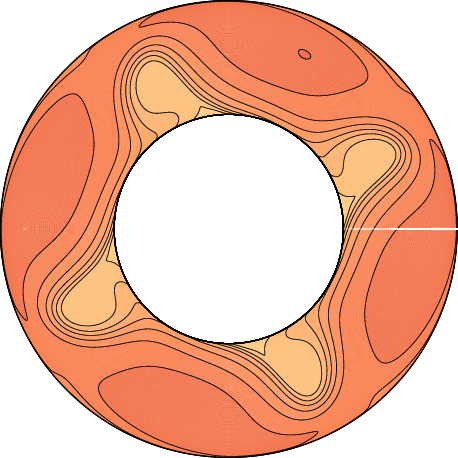}
    \includegraphics[width=0.13\linewidth]{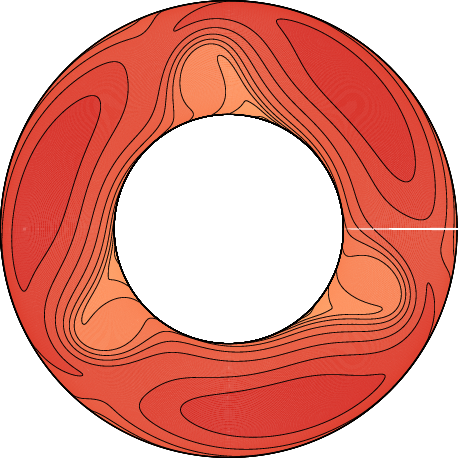}\\
    \includegraphics[width=0.13\linewidth]{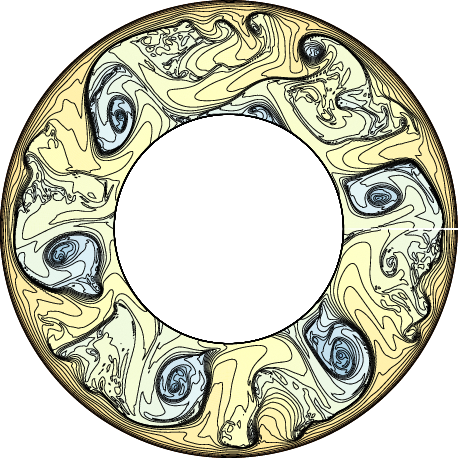}
    \includegraphics[width=0.13\linewidth]{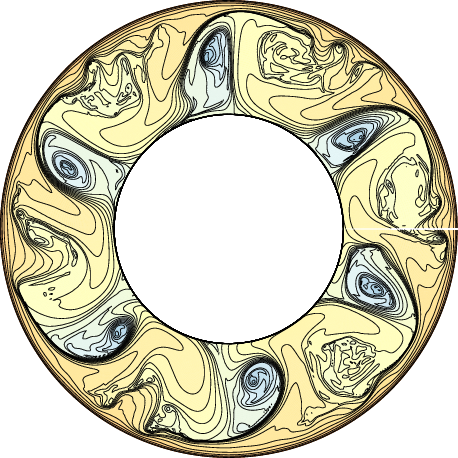}
    \includegraphics[width=0.13\linewidth]{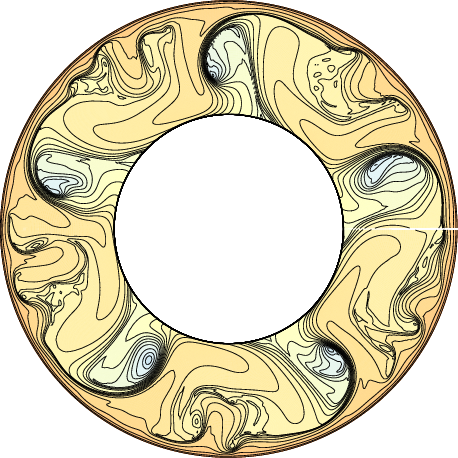}
    \includegraphics[width=0.13\linewidth]{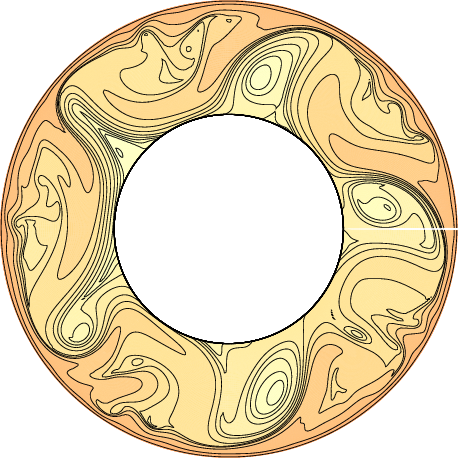}
    \includegraphics[width=0.13\linewidth]{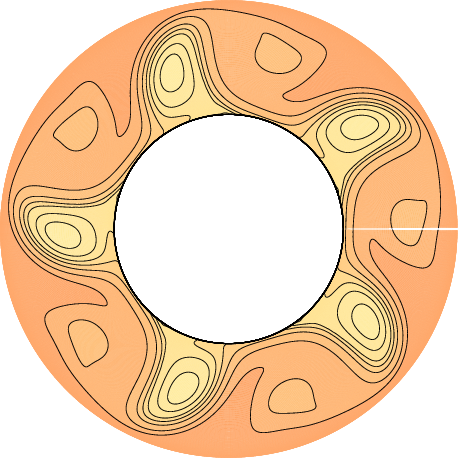}
    \includegraphics[width=0.13\linewidth]{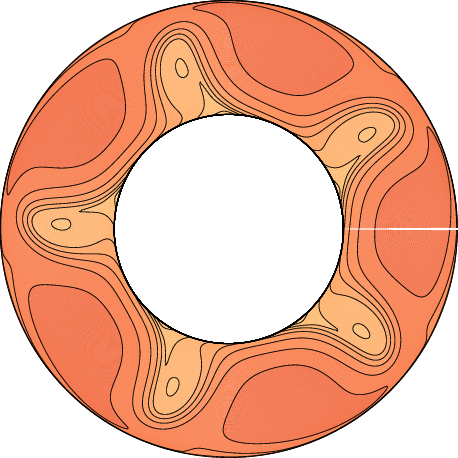}
    \includegraphics[width=0.13\linewidth]{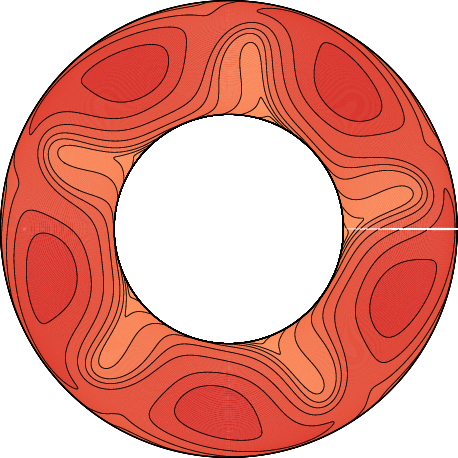}
    \caption{Snapshots of the temperature surface highlighting the influence of the quiescent temperature on the baroclinic instability features. From left to right, the quiescent temperature  systematically increased in increments of $1$  within the interval $-3 \le \Theta_q\le +3$. From top to bottom, the rotation rates are incrementally raised by $0.1$, spanning from $0.3 \le \Omega \le 0.6 \,rpm$.\label{fig:snapshot}}
  \end{figure*}

In this section, we investigate the characteristics of baroclinic waves when we vary the rotation rate $\Omega$ while maintaining constant values for the temperature difference and the quiescent temperature, denoted as $(\Delta T, T_q)$, following a setup similar to the atmospheric BT experiment presented by \citet{rodda2020new}. Specifically, we fix the temperature difference at $\Delta T = 2.5 K$, and the rotation rate $\Omega$ is varied within the range of $0.3-0.6\,rpm$.
We expect the onset of the baroclinic waves to occur in this range of parameters, as indicated by previous studies \citep{read2020baroclinic}.
The values of the quiescent temperature $T_q$ are selected to align with practical experimental conditions, falling within the range of $\pm3\Delta T$.
It is important to note that altering the $T_q$ parameter is analogous to modifying the heat flux at the free surface, with the corresponding dimensionless temperature range denoted as $-3\leq \Theta_q\leq 3$. 
Negative values of $\Theta_q$ indicate a colder surrounding environment relative to the surface temperature, resulting in surface cooling. Conversely, positive values of $\Theta_q$ correspond to warming up the water at the free surface of the BT.
Figure~\ref{fig:snapshot} shows the surface temperature of the BT for different values of quiescent temperature and rotation rate. We highlight qualitative similarity between structures observed in certain numerical results and those obtained in BT experiments, including simpler configurations such as the one illustrated in Figure~\ref{fig:simpleBaroclinic}, especially considering lower values of quiescent temperature and higher rotation rates (as we move to the bottom left case in Figure~\ref{fig:snapshot}).

Figure~\ref{fig:snapshot} reveals two noteworthy qualitative observations. Firstly, the stability of the system is intricately linked to the quiescent temperature, with lower temperatures favoring the transition process. Specifically, when the temperature is lowered to $\Theta_q=-3$ (left column in Fig.~\ref{fig:snapshot}), a rapid transition from the axisymmetric regimes occurs, even under the influence of the low rotation rates within the range of $0.3-0.4\,rpm$. Conversely, elevating the surface temperature to $\Theta_q=+3$ leads to the patterns observed in the right-hand column of Fig.~\ref{fig:snapshot}, indicating a considerably delayed transition, starting within the range of $0.4-0.5\,rpm$. Moreover, we note that the development or inhibition of small-scale features is also influenced by the quiescent temperature or, equivalently, cooling or heating of the baroclinic cavity free-surface. While the influence of temperature on small-scale features has been mentioned in the literature \citep{randriamampianina_2013,randriamampianina_2015,vLarcher,rodda2020new}, to the best of our knowledge, no systematic study has previously shown this dependence of small-scale feature development on the external room temperature, as we will discuss here.

\subsection{Control of the thermal stratification}
\begin{figure}[htbp!]
 \begin{subfigure}{\textwidth}
    \centering   
    \caption{$\Theta_q=-3$}
    \includegraphics[width=0.80\linewidth]{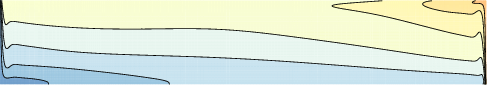}

\end{subfigure}
\begin{subfigure}{\textwidth}
\centering
\caption{$\Theta_q=+3$}
    \includegraphics[width=.80\linewidth]{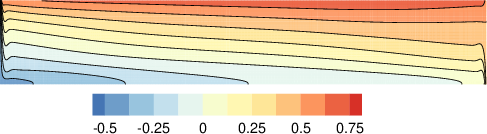}
    
\end{subfigure}
   
   \caption{Contour maps of the mean temperature $\overline{\Theta}$ at a rotation rate of $\Omega=0.5\, \text{rpm}$ with a dimensionless quiescent temperature (a)~$\Theta_q=-3$, and (b)~$\Theta_q=+3$, both presenting 11 contour levels ranging from $-0.5$ to $0.75$.
   \label{fig:stratification}}
 \end{figure}

\begin{figure}[htbp!]
   \begin{center}
     \includegraphics[height=7cm]{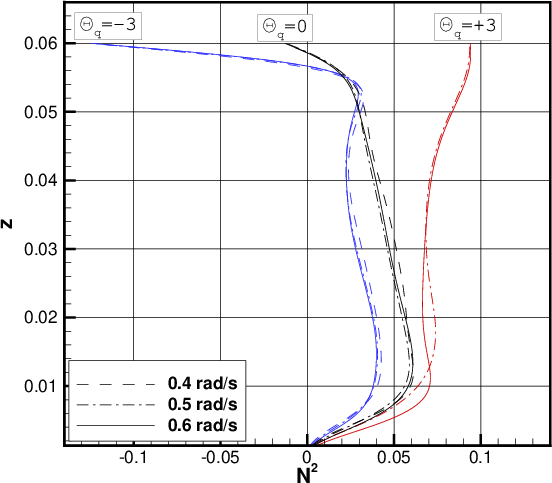}
   \end{center}
   \caption{Vertical profiles of the Brunt–V\"ais\"al\"a frequency $N^2$ at mid-radius for different rotation rates $\Omega$ and quiescent temperatures $\Theta_q$.\label{fig:Brunt_profiles}
   }
 \end{figure}

In this section, we outline how the quiescent temperature $\Theta_q$  has an effect on the global thermal stratification and the underlying baroclinic wave dynamics.
A typical thermal stratification of the BT in our experiments is presented in Figure~\ref{fig:stratification}, where two contour maps show the mean temperature $\overline{\Theta}(r,z)$ in the radial-axial direction, averaged both in time and space (in the azimuthal direction). 
The top picture in Figure~\ref{fig:stratification} shows a temperature contour map for a tank rotating at $\Omega=0.5\,\textrm{rpm}$, with a constant dimensionless quiescent temperature $\Theta_q=-3$. 
This implies that the reference temperature $(T_c+T_h)/2$ at the surface is warmer than the ambient (cold) temperature, resulting in a configuration that resembles an adiabatic boundary condition. The bottom map in Figure~\ref{fig:stratification} shows a warm environment compared to the fluid surface, where we note that, outside the thermal boundary layers, there is a smooth decrease of the thermal stratification in the vertical direction, with an almost uniform temperature in the central region of the cavity.
This analysis is supported by Figure~\ref{fig:Brunt_profiles}, which shows the averaged Brunt–V\"ais\"al\"a frequency $\left(N^2=\frac{g}{\overline{\Theta}}\partial_z\Theta \right)$ vertical profiles  for different rotation rates $\Omega$ and quiescent temperatures $\Theta_q$. The radial position is fixed at mid-gap.
When we examine the relationship between $N$ and the rotation rate ($\Omega$) in Figure~\ref{fig:Brunt_profiles}, we observe that variations in $\Omega$ have a very small impact on the $N^2$ profiles. In contrast, changes in $\Theta_q$ have a larger influence on $N^2$ profiles. Despite the differences produced by changes in $\Theta_q$, all the curves show an almost uniform vertical stratification at mid-height. 
The deviations become larger as we approach the Ekman layer. Although both upper and lower boundary regions exhibit higher values of $|\partial N^2/\partial z|$ compared to those observed near the mid-height, an asymmetry between the two boundary regions is evident.

 In the lower Ekman layer, variations in the Brunt-V\"ais\"al\"a frequency indicate that the boundary layer responds similarly to changes in the control parameters $(\Omega, \Theta_q)$.
In contrast, the values of  $|\partial N^2/\partial z|$ closer to the free surface behave differently depending on the quiescent temperature.
Furthermore, we notice that for lower values of the quiescent temperature, we observe imaginary values of the Brunt-V\"ais\"al\"a frequency ($N^2<0$). This could explain the occurrence of small scale features at the free surface. If $N$ becomes complex, it indicates that the stratified fluid is unstable with respect to small-scale perturbations, i.e., when $N^2$ is negative, the buoyancy force and the stratification are opposing each other, which could lead to convective instability if a critical value of the vertical temperature profile is reached. 
\subsection{Effects on the vertical temperature profiles}

It is well known that the baroclinic instability perturbs the mean vertical temperature gradient.
This is generally described using the pair of control parameters $(Ta,Bu)$ representation of the regime \citep{oNiel1969stability,rodda2020new}.
Based on the work of \citet{oNiel1969stability}, we propose in this section a novel pair of parameters leading to an alternative representation of the marginal stability curve. 
We will then investigate the sensitivity of the axial (vertical) temperature profiles to the quiescent temperature considering a re-scaling that couples the vertical ($\Delta_VT$) and the horizontal ($\Delta_HT$) temperature gradients into one more generalized marginal instability curve. 
In other words, we will look at the results considering a parametrization that does not have an explicit dependency on each temperature gradient independently, but on their ratio $\epsilon_b=\Delta_HT/\Delta_VT$.

According to \citet{hide1967theory}, the effective vertical ($\Delta_V T$) and horizontal ($\Delta_HT$) mean temperatures can be related to the prescribed horizontal temperature difference  :
\begin{equation} \label{eq:sigma_r}
  \sigma_r = \int^{\eta_o}_{\eta_i} \int^{\Gamma}_{0} \frac{\partial}{\partial r} \overline{\Theta} \mathrm{d} r \mathrm{d} z,\quad 
  \sigma_z = \int^{\eta_o}_{\eta_i} \int^{\Gamma}_{0} \frac{\partial}{\partial z} \overline{\Theta} \mathrm{d} r \mathrm{d} z 
\end{equation} 
where, as a reminder, $\Theta=(T-T_r)/\Delta T$ is the non-dimensional temperature, $T_r$ is the reference temperature, and the bar above the variables ($\overline{\cdot}$) denotes time and azimuthal space averaged quantities.
These parameters measure a normalised average temperature gradient that reflects the vertical temperature field's deviation from the prescribed horizontal temperature $\Delta T$, taking into account the relations $\Delta T=T_h-T_c$ by defining the  $\Delta_VT=\sigma_z \Delta T$ and $\Delta_HT=\sigma_r \Delta T$.

\begin{figure}[htbp!]
  \begin{center}
    \includegraphics[height=7cm]{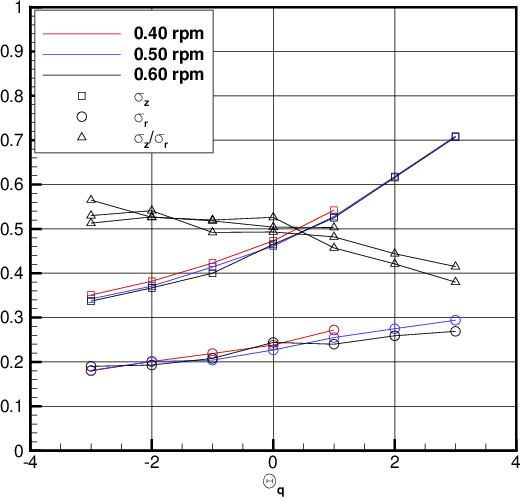}
  \end{center}
  \caption{Mean radial $\sigma_r$, axial $\sigma_z$, and $\epsilon_b=\sigma_r/\sigma_z$, versus the quiescent temperature for different rotation rates. \label{fig:normalized_grad}}
\end{figure}

Figure~\ref{fig:normalized_grad} presents $\sigma_r$ and $\sigma_z$ versus the quiescent temperature for different rotation rates.
The $\sigma_z$ values were found to be sensitive to the quiescent temperature, and less dependent on the rotation rate.
As expected, the strongest thermal stratification occurs for the warmest quiescent temperature $\Theta_q=3$.
The sensitivity is not negligible, and cooling or heating the free surface has different influences on the axial temperature profiles,  given that the value of $\sigma_z$ is twice as small as that observed for the cooling case ($\Theta_q=-3$) when compared to the case $\Theta_q=+3$, with a heated surface. 

An obvious consequence of the changes in $\sigma_z$ seen in Figure~\ref{fig:normalized_grad} is the changes in the onset of baroclinic instability, that depends on the Burger number, $ \textrm{Bu}=\sigma_z H g \alpha \Delta T/ (b-a)^2$. The quantity $\sigma_z$ is also related to the Burger number since it can be redefined as \citep{borchert2014gravity}:
\begin{equation} \label{eq:Bu_sigma_z}
  Bu = \sigma_z Ro_T . 
\end{equation}
 We can then notice that, for the BT configuration considered, the Burger number could be heavily affected by the heat transfer at the free surface.
Such a dependence was noticed earlier in the work of \citet{oNiel1969stability}, who proposed that the stability criterion for the development of the baroclinic instability relies on $\sigma_z$ through the Burger number, as well as on the ratio $\epsilon_b = \sigma_z / \sigma_r$, also presented in Figure~\ref{fig:normalized_grad}. In other words, the stability criterion varies with both the horizontal and vertical temperature differences. This re-scaling leads to one curve that depends only on the values of $\epsilon_b$.  
Nevertheless, \citet{oNiel1969stability}'s neutral stability equation can also be re-scaled like $\textrm{Ta}\epsilon^{8/3}_b$, suggesting that we could then eliminate the dependency on $\sigma_r$ (see Appendix~\ref{Appendix:oNielStability}).
\begin{figure}[htbp!]
\begin{center}
  \includegraphics[width=0.8\linewidth]{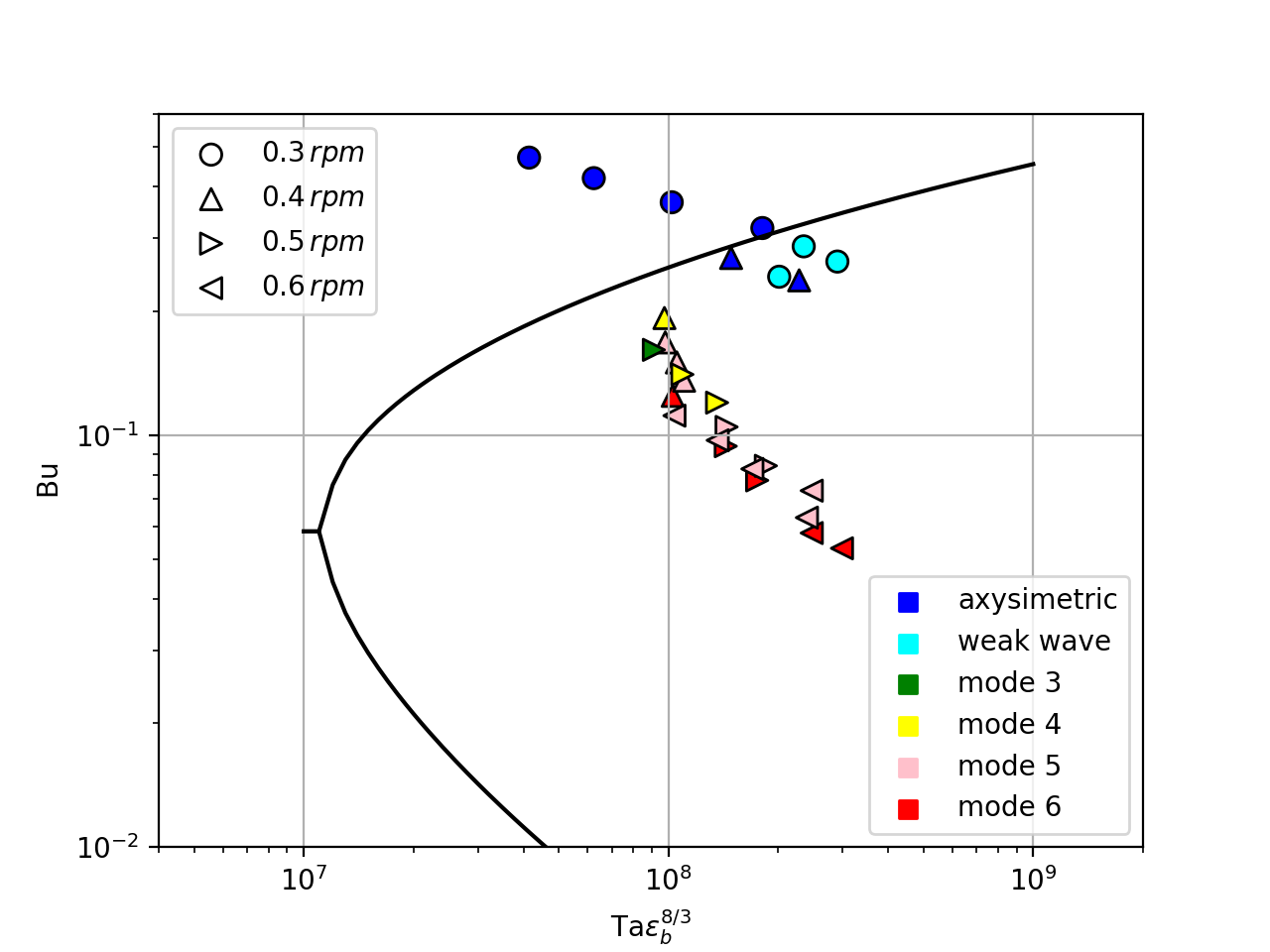}
  \caption{ Marginal stability diagram for the atmosphere-like regime on a BT confronted with the results from numerical simulations. \label{fig:oneil_curve}  }
  \end{center}
\end{figure}
Figure~\ref{fig:oneil_curve} presents O'Neil's
stability diagram with this new scaling and confronts it with the results of our numerical simulations. Based on this re-scaling, the onset of baroclinic instability could be approximately predicted by our numerical results, with all the axisymmetric and weak wave regimes observed above or, at least, near the marginal curve. Simulations leading to steady waves or irregular regimes can be found below the marginal stability curve. All of these results are in accordance with the ideas presented by \citet{oNiel1969stability} and the conclusions of \citet{douglas1973thermal}, given this alternative scaling for a better description of the instability onset.

\subsection{Effects on drift rate and thermal variability.}

\begin{figure}[htbp!]
\begin{center}
  \includegraphics[width=0.45\textwidth]{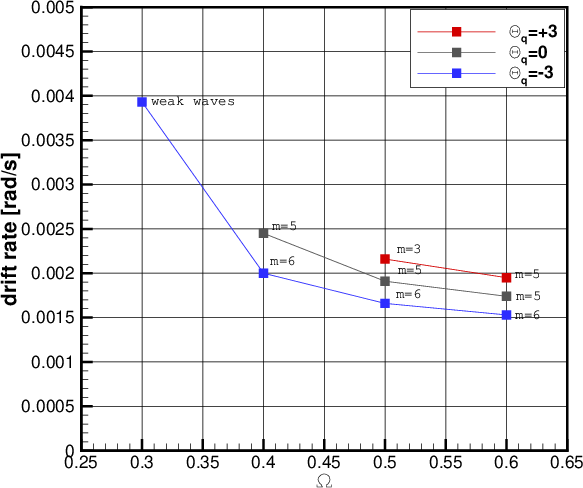}
  \caption{ The drift rate $c(t)$ at a mid radial position and at the top free surface as a function of the rotation rate $\Omega$.  The smaller rotation rates for $\Theta_q=0,3$ are not shown because the instability is not developed for those cases. \label{fig:drift}}
  \end{center}
\end{figure}

Drift rate and thermal variability are closely linked to baroclinic waves, making it essential to carefully consider how they are influenced by the parameters previously presented. Figure~\ref{fig:drift} illustrates the drift velocity $c(t)$ as a function of the rotation rate $\Omega$ for different quiescent temperatures, at a mid radial position and at the top free surface. The dominant wavenumbers are given in the plot. Regardless of the quiescent temperature, the drift decreases with the rotation rate, which is consistent with observations from small-tank experiments and simulations (Fig. \ref{fig:drift_vs_omega_small_tank}). Note that, in the simulations considered here, the parameters of the big tank configuration are used, so that a larger surface is considered.
Furthermore, for a constant rotation rate, a higher quiescent temperature results in a faster drift, as expected since it corresponds to larger Burger numbers.

\begin{figure}[htbp!]
\begin{center}
  \includegraphics[width=0.45\textwidth]{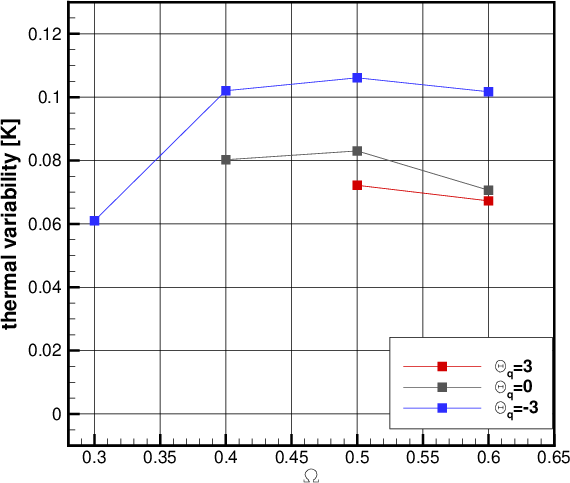}
  \caption{ Mean thermal variability $\overline{\sigma_z}$ for different quiescent temperature values as a function of the rotation rate $\Omega$. The smaller rotation rates for $\Theta_q=0,3$ are not shown because the instability is not developed, therefore, $\overline{\sigma_z}=0$. \label{fig:sigma}}
  \end{center}
\end{figure}

\begin{figure}[htbp!]
{
\centering
\begin{subfigure}{\textwidth}
\centering
\caption{ $\theta_q=-3$ }
    \includegraphics[width=0.7\linewidth]{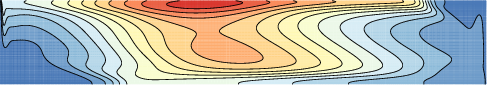}
    
\end{subfigure}
\begin{subfigure}{\textwidth}
\centering
\caption{ $\theta_q=0$ }
    \includegraphics[width=0.7\linewidth]{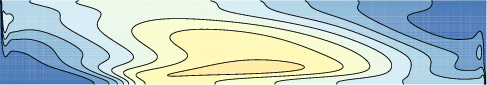}
  
\end{subfigure}
\begin{subfigure}{\textwidth}
\centering
\caption{ $\theta_q=+3$ }
    \includegraphics[width=0.7\linewidth]{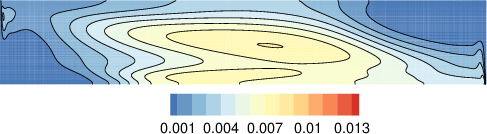}
    
\end{subfigure}
  \caption{Isocontours of the {\it rms}-temperature in the radial-axial plane, averaged over time and azimuthal direction for a rotation rate fixed at $0.6\,rpm$, 13 levels . \label{fig:rms_temperture}}
  }
\end{figure}

The thermal variability $\overline{\sigma_z}$ variations with respect to the rotation rate are shown in Figure~\ref{fig:sigma} for three different quiescent temperatures: $\Theta_q=-3,0,3$. 
The changes in $\sigma_z$ with respect to rotation are related to the instability's dominant wavenumbers. This comes from the fact that, independently of the quiescent temperature influence, the drift velocity decreases with the rotation rate, which agrees with the findings from the small-tank experiments and with the numerical simulations presented on the left side of Table~\ref{tab:Params_BBT}, performed by \cite{rodda2020new} (see Figure~\ref{fig:drift_vs_omega_small_tank} and \ref{fig:sigma_vs_omega}).
It is also clear in Figure~\ref{fig:sigma} that $\overline{\sigma_z}$ is significantly influenced by the quiescent temperature $\Theta_q$. Notably, temperature fluctuations at the free surface are approximately twice as high for a cooling quiescent temperature, related to negative values of $\Theta_q$. This observation should be interpreted in the light of the negative values of $N^2$ observed previously in the upper Ekman boundary layer (see Figure~\ref{fig:Brunt_profiles}), which can lead to a Rayleigh-Taylor-type instability near the surface, that would increase the level of surface temperature fluctuations.
%
%
%
%
%
Figure~\ref{fig:snapshot} shows snapshots of the surface temperature,  revealing the appearance of these small-scale structures when we move towards small quiescent temperature values and high rotation rates (towards the bottom-left of the figure). These small-scale patterns observed at the free surface can be associated with higher root mean square (rms) temperature values, as shown in Fig.~\ref{fig:rms_temperture}. The figure presents a comparison of \textit{rms} temperature values between $\Theta_q=-3$ (top figure) and $\Theta_q=+3$ (bottom figure) in the radial-axial cross-section. The temperature fluctuations presented are averaged in time and azimuthal direction. 
From Fig.~\ref{fig:rms_temperture}, it is clear that the larger fluctuations are found directly below the free surface for the case $\Theta_q=-3$. For the cases $\Theta_q=0$ and $\Theta_q=+3$, this maximum is smaller and, most importantly, it is found at the bottom region of the cavity.
This underlines, once again, the influence of the heat transfer at the free surface,  which appears to enhance temperature fluctuations associated with the presence of the small-scale features. 
It is important to highlight that, at this stage, qualitatively, there seems to be a relation between the small scales observed and their impacts on the baroclinic wave dynamics, but we are not yet able to quantify this statement.

\begin{figure}[htbp!]
\begin{center}
    
  \includegraphics[width=0.19\textwidth]{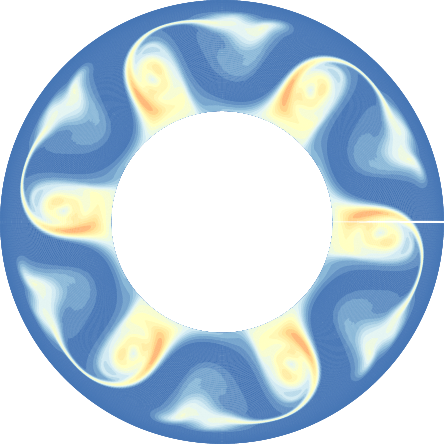}
  \includegraphics[width=0.19\textwidth]{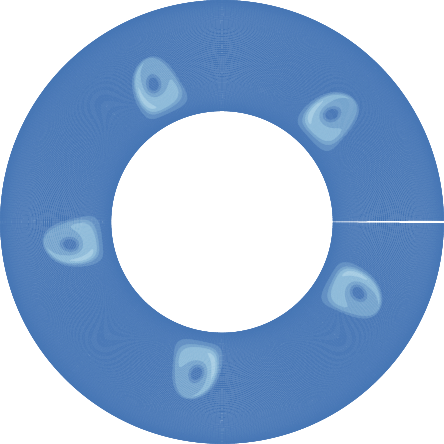}
  \includegraphics[width=0.19\textwidth]{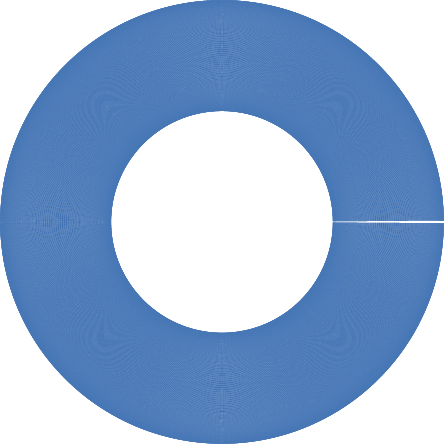}\\
  \includegraphics[width=0.19\textwidth]{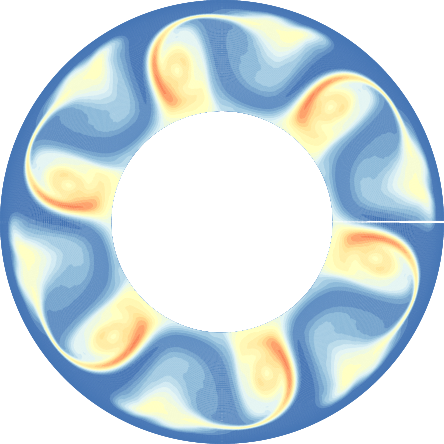}
  \includegraphics[width=0.19\textwidth]{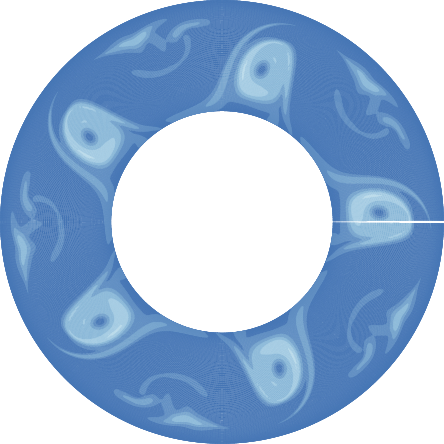}
  \includegraphics[width=0.19\textwidth]{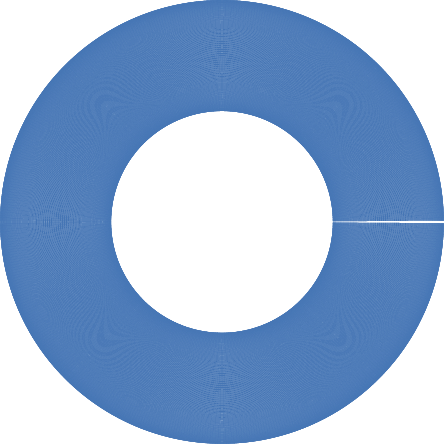}\\
  \includegraphics[width=0.19\textwidth]{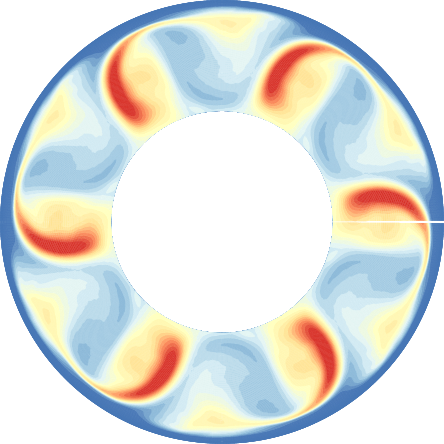}
  \includegraphics[width=0.19\textwidth]{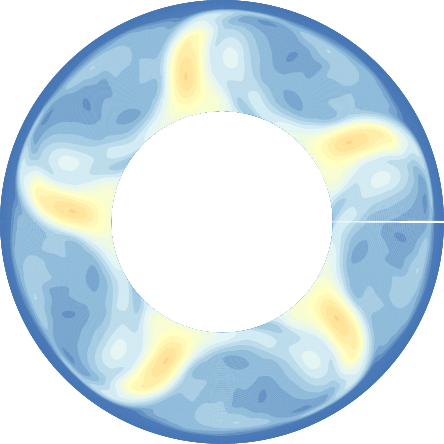}
  \includegraphics[width=0.19\textwidth]{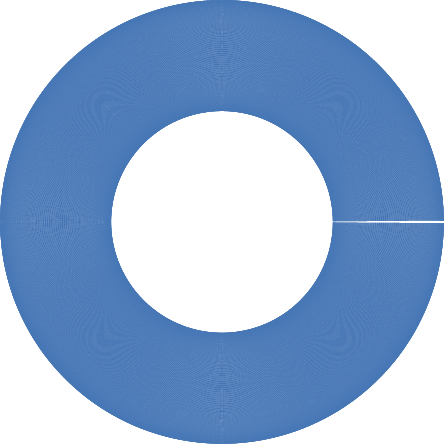}
  \end{center}
    \caption{ Time mean \textit{rms} temperature at the surface of a BT for varying temperatures $\Theta_q=-3,0,+3$ (from left to right) and rotation rates of $0.4$, $0.5$ and $0.6\,rpm$ (from top to bottom). The figures are shown in a frame that rotates with the drift velocity of the baroclinic wave. \label{fig:snapshots3}}
  \end{figure}

We now analyzed the temperature behavior at the free surface of a BT flow under different rotation rates and quiescent temperatures regarding the large instability scales. 
In Figure
~\ref{fig:snapshots3}, we present the time-mean
root mean square (RMS) values of the temperature at the reference frame of the wave-drift, i.e., if we look at the flow from the perspective of an observer that is moving at the drift velocity. 
The Figure illustrates how the rotation rate and quiescent temperature influence the growth of larger flow structures. Notably, we observe significant differences for specific combinations of rotation rate ($\Omega$) and quiescent temperature ($\Theta_q$), where small-scale structure influences are observed.
Note that, time-averaged temperatures (not shown) closely resemble the instantaneous temperature fields presented in Figure~\ref{fig:snapshot}, particularly for situations with low rotation rates and higher ambient temperatures (towards the top-right figures), i.e., those cases where we do not observe many small-scale structures.
Moreover, in scenarios such as ($\Omega=0.5 \text{ rpm}, \Theta_q = -3$), ($\Omega=0.6 \text{ rpm}, \Theta_q = -3$), or ($\Omega=0.6 \text{ rpm}, \Theta_q = 0$) (towards the bottom left corner in Fig. \ref{fig:snapshots3}), we observe a higher background fluctuation between two waves compared to other cases. This hints at the possibility of convective instabilities near the free surface.  
Additionally, the maximum fluctuation remains located at the front of the baroclinic waves, suggesting a potential connection with baroclinic wave vacillations.  
Interestingly, fluctuations along the fronts of a baroclinic instability were observed experimentally by~\cite{rodda2020new}, associated with gravity waves emission.

%
These latest findings are consistent with our prior analysis, demonstrating that decreased quiescent temperatures correlate with a reduced Burger number, giving rise to an amplitude oscillation-like regime. This insight illuminates the intricate relationship between rotation rate, quiescent temperature, and the emergence of different instability patterns within the BT flow.

\begin{figure}[htbp!]
\begin{center} 
  \includegraphics[width=0.6\textwidth]{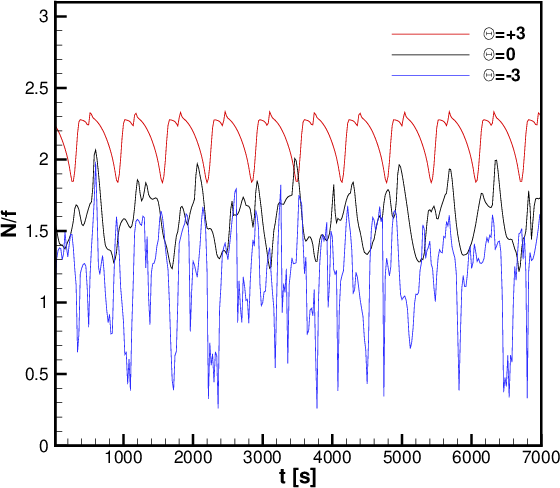} 
  \end{center}
    \caption{ Time series of the parameter $N/f$ for varying quiescent temperatures $\Theta_q=-3,0,+3$ and for the rotation rate $0.6\,rpm$.  \label{fig:N_f}}
  \end{figure}

Figure~\ref{fig:N_f} shows three distinct time series illustrating the $N/f$ parameter under quiescent temperatures of $\Theta_q=-3,0,+3$ for a rotation rate of $\Omega=0.6 \text{ rpm}$, highlighting the impact of ambient temperature on the observed trends. A warmer environment compared to the fluid surface tends to dampen amplitude variations in $N/f$, while a colder environment leads to the opposite effect. Our simulations reveal that fluctuations considering a cold environment become more asymmetric relative to their mean value, akin to observations made experimentally presented in Figure~\ref{fig:bigtank_rodda}.
Additionally, all simulations employing adiabatic boundary conditions fail to accurately capture the increased amplitude variations in $N/f$, as discussed in Section~\ref{subsec:Validation}. However, when these conditions are revised to include heat flux through the surface, the simulation results align more closely with the experimental results presented in Figure~\ref{fig:bigtank_rodda}. This highlights the significance of incorporating heat flux in the simulation of baroclinic instabilities. Such inclusion not only substantially reduces the disparities between experimental and numerical outcomes observed with adiabatic boundary conditions, but also introduces distinct levels of turbulent features to the problem.

\section{Analysis of the small scale features based on Harmonic decomposition.}\label{sec:small-scales}

\begin{figure}[htbp!]
  \includegraphics[width=0.49\linewidth]{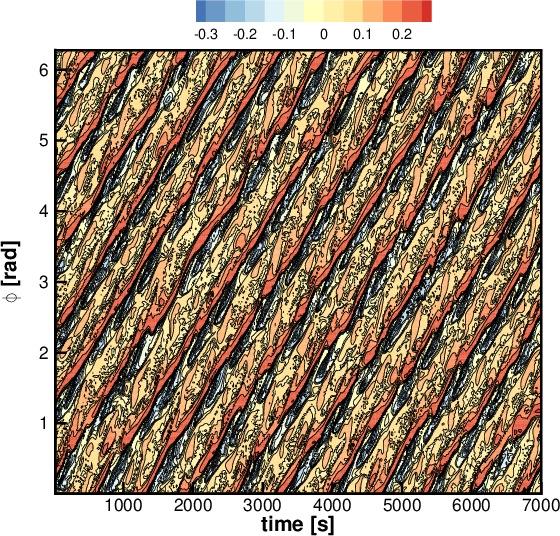}
  \includegraphics[width=0.49\linewidth]{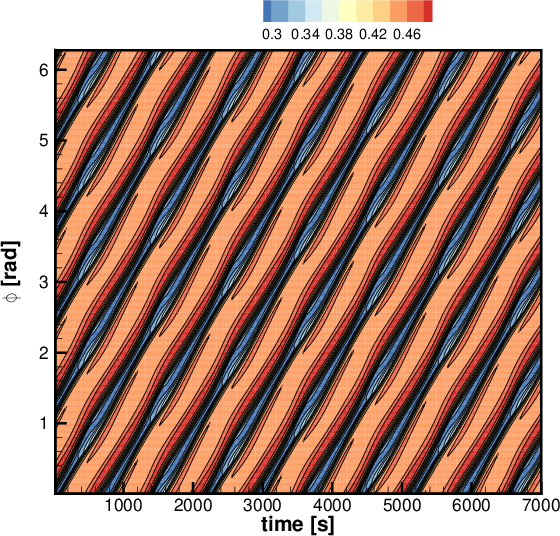}
   \caption{Space-time (Hovm\"oller) representation of the azimuthal velocity, in the drift frame of reference. The radial and axial positions are fixed at their midpoints between the inner and outer boundaries. The x-axis represents time, and the y-axis represents the azimuthal direction. The BT rotates at a constant angular velocity of $0.6$ rpm. The plot on the left shows results for $\Theta_q = -3$, i.e., for a warm surface that is cooled down by the surrounding environment. The right plot shows data for $\Theta_q = +3$, i.e., a cold surface that is warmed up by the surrounding environment.
     }\label{fig:Hovemoller}
\end{figure}

\begin{figure}[htbp!]
\includegraphics[width=0.49\linewidth]{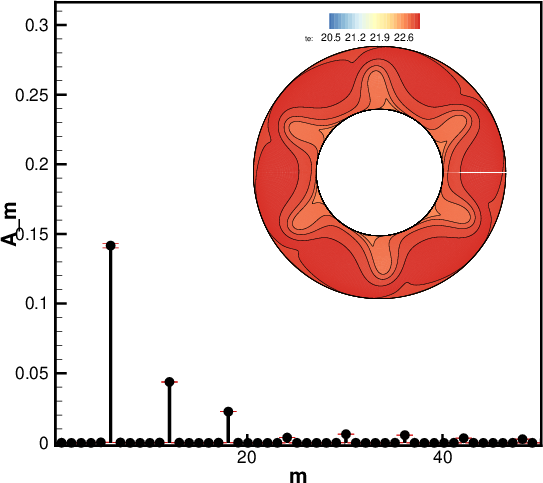}
\includegraphics[width=0.49\linewidth]{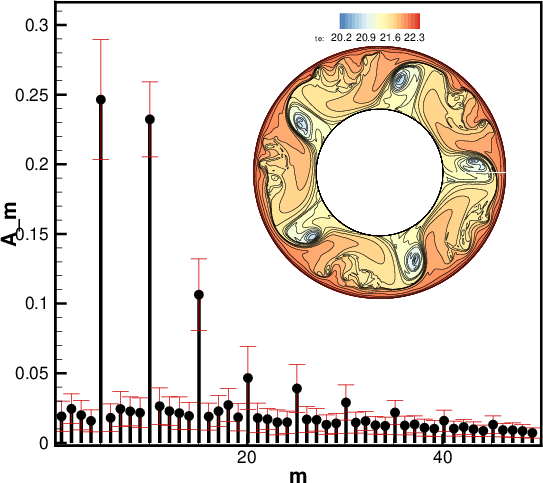}
\caption{Fourier coefficient amplitudes in the azimuthal space direction derived from time-averaged RMS temperatures. The BT rotation rate is fixed at $0.3\, rpm$, and we observe cases of heating ($\Theta_q = +3$, on the left) and cooling ($\Theta_q = -3$, on the right) the BT surface. The space Fourier spectra are taken at a fixed radial position at the mid-gap cavity at the free surface of the cavity. Error bars considering the temperature fluctuations in time are also displayed in both (left and right) spectra. Snapshots of the RMS temperature for each corresponding case are displayed at the top right of the spectra. } \label{fig:FourierSpectra}
\end{figure}

To isolate large-scale dynamics from the small scales, we use a statistical method that works as a low-pass filter by averaging the results obtained at each rotation frame rate. This procedure is detailed in the appendix~\ref{Appendix:StatisticalAnalysis}. This approach allows us to focus on the distinct behavior of the flow components, such as the depiction of the baroclinic dynamics by means of the drift related to the thermal variability.
Figure~\ref{fig:Hovemoller} presents such an analysis using space-time (Hovm\"oller) plots with azimuthal y-direction and temporal x-direction. The data have been taken from fixed radial and axial positions at their mid-distance between inner and outer (top and bottom) boundaries. 
The diagrams are presented for a fixed angular velocity of $0.6\,rpm$ and two quiescent temperatures : $\Theta_q=-3$ (on the left) and $\Theta_q=1$ (on the right).
Note that using the average in the drift frame helps distinguish fluctuations associated with large-scale oscillations from those linked to small structures. Both scenarios show regular large-scale baroclinic oscillations. However, for $\Theta_q=-3$, which corresponds to a "warm" interface being cooled by the environment (as seen on the left side of Fig.~\ref{fig:Hovemoller}), small-scale features with a more random spatial pattern develop between the baroclinic large-scale fluctuations. The space-time diagrams also highlight the coherence of these fluctuations.
Figure~\ref{fig:FourierSpectra} shows the Fourier coefficient amplitudes of the RMS temperature space spectra after applying this time-averaging technique. The cases correspond to heating ($\Theta_q = +3$) and cooling ($\Theta_q = -3$) the BT surface, with a fixed rotation rate of $0.3\, rpm$, respectively in the left and right figures. Representative snapshots of each case are inserted in the top right part of the figures. Error bars sized by their standard deviations are also plotted to highlight vacillating or irregular regimes. 
Note that, in the context of cooling at the surface resulting in $N^2<0$, as we discussed earlier, a significant increase in error bars is notable as the quiescent temperature decreases. Additionally, when examining the snapshots of RMS temperature in Figure~\ref{fig:FourierSpectra}, we observe elevated background fluctuation levels in the non-dominant modes for the cooling surface scenario (right side plot of Figure~\ref{fig:FourierSpectra}). This is reflected in the increased intensity of the small-scale signature that we observe in the spectrum. The signature of the dominant modes could be attenuated by analyzing the flow at the drift frame, i.e., if we look at the flow from the perspective of an observer that is moving at the drift velocity.

\begin{figure}[htbp!]
  \includegraphics[width=0.31\textwidth]{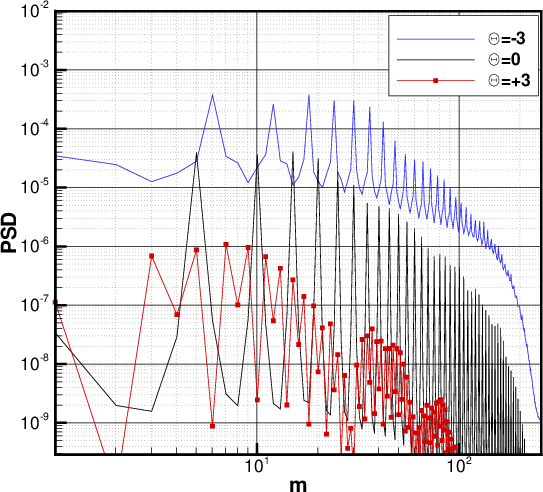}
  \includegraphics[width=0.31\textwidth]{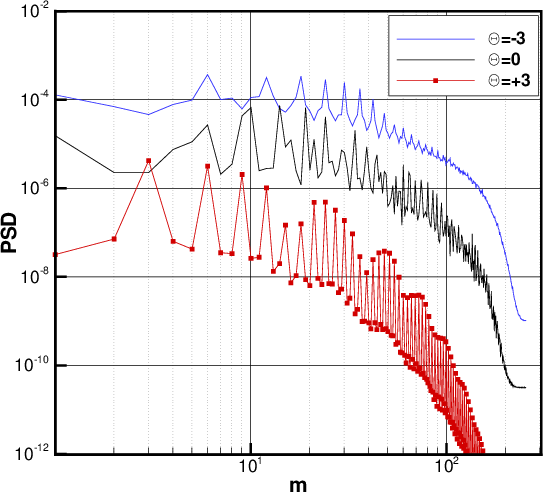} 
  \includegraphics[width=0.31\textwidth]{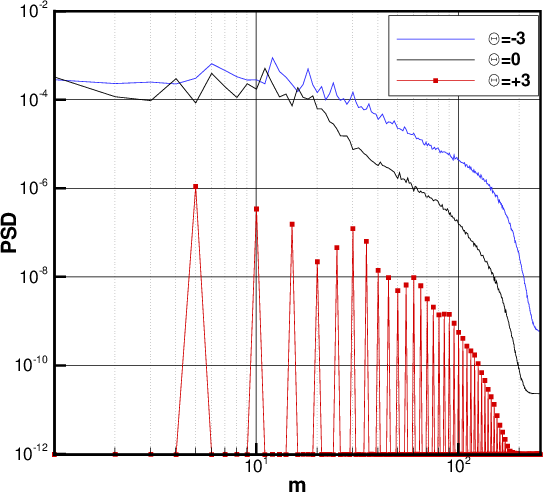} 
  \caption{ Azimuthal Fourier spectra for rotation rates of (from left to right) $0.4\,rpm$, $0.5\,rpm$  and $0.6\,rpm$.} \label{fig:largeScaleSpectra}
\end{figure}

In Figure~\ref{fig:largeScaleSpectra}, the space spectra for different $\Theta_q$ and rotation rates are presented, showing the transition from a discrete spectrum for a rotation rate of $0.4\,rpm$ (on the left)  towards a spectrum without peaks for higher rotation rates (on the right). This smoother decaying spectra occur either from a high rotational speed, or smaller quiescent temperature (cold environment compared to the surface).
Towards the rotation speed of $0.6\,rpm$ and the quiescent temperature $\Theta_q=-3$ (right side of Figure~\ref{fig:largeScaleSpectra}), the spectra observed have almost no peaks and present a decay slope of the order of $-8/3$. 
The slope and the smooth decaying behavior of the spectra might suggest that we are moving toward the transition to turbulence, but this result would have to be confirmed by a numerical simulation with a higher rotation rate, which was not performed up to this point.

\newpage

\section{Conclusions}\label{sec:conclusions}
In this study, we performed an analysis of a baroclinic instability flow within a cylindrical baroclinic tank. We carefully examined the effects of two critical factors: the rotation rate ($\Omega$) and the quiescent temperature ($\Theta_q$) on multiple aspects of the flow behavior. Our exploration provided insights that clarify the complex relationship between these variables. These insights can explain differences observed between numerical simulations and laboratory experiments since it became evident from our simulations that the heat exchange at the surface of baroclinic tanks plays an important role on the development and the characteristics of baroclinic instabilities. 

After showing the qualitative impacts of $\Theta_q$ and $\Omega$ on the baroclinic instability, we explored their influence on the global thermal stratification of the flow. Our findings highlighted that $\Theta_q$ plays a crucial role in determining the thermal stratification near the free surface. Cooling the upper surface, leading to negative $\Theta_q$, resulted in negative Brunt-Väisälä frequency ($N^2$) values in the upper Ekman boundary layer, leading to the emergence of small-scale structures at the free surface. On the other hand, heating the surface led to a more stable stratification near the surface.
This, in turn, affected the Burger number, which is a critical parameter for the onset of baroclinic instability. We observed further that, in contrast to $\Theta_q$ variations, changing $\Omega$ had minimal effects on the $N^2$ profiles.
Drift rate 
and thermal variability
were also found to be closely connected to the baroclinic waves, with a clear trend that higher $\Theta_q$ values increase the drift rate.

To isolate large-scale dynamics from small scales, we employed a statistical method that acts as a low-pass filter. This analysis revealed that, for a fixed rotation rate, cooling the upper surface results in a higher background fluctuation, indicating convective feature development. The evaluation of different azimuthal Fourier spectra further suggested that, as $\Omega$ increased or $\Theta_q$ decreased, a transition to spectra without clear peaks occurs, which might characterise an evolution to turbulent regimes. Future work should explore higher rotation rates and their effects on the transition to turbulence in such systems, as well as a more extensive confrontation of our numerical results with future laboratory experiments.\\

\noindent \textbf{Acknowledgement:} UH acknowledges support from the DAAD project ``Combined studies of baroclinic waves with methods of data assimilation'' (57560889). UH thanks R. St{\"o}bel, S. Rohark, and S. Misera for technical support. We thank P. Szabo and Y. Sliavin for proofreading.

\newpage
\appendix
\section{Statistical analysis in the {\it drift frame}}\label{Appendix:StatisticalAnalysis}
\begin{figure}[htbp!]
  \includegraphics[width=0.32\linewidth]{figures/png/snapshot-0p6_c7p5.png}
  \includegraphics[width=0.32\linewidth]{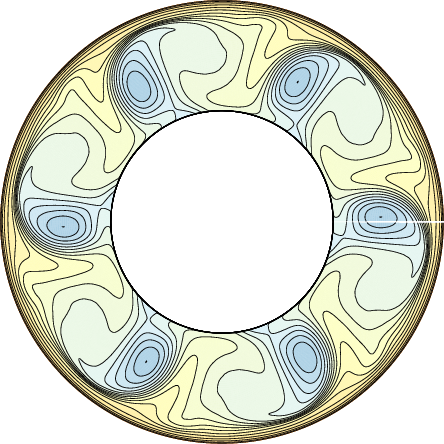}
  \includegraphics[width=0.32\linewidth]{figures/gallery/rms_0p6_7p5.png}
  \caption{Example of surface temperature data and its statistics in the drift frame. Left: snapshot at a given time; center: mean; right: root mean square (rms). \label{fig:StatisticsDriftFrame}}
\end{figure}

Baroclinic flow is characterised by large-scale dynamics that can be specifically separated to identify the small scales.
As an example, we can mention the three-term decomposition introduced by \cite{Hussain_Reynolds_1970}.
Here we propose post-processing our data by following the baroclinic waves in their rotation.
In this new frame, the temperature and velocity fields are obtained read:
\begin{equation}
  \phi^{d} \left(  t , r , \theta \right) = \phi \left( t ,  r , \theta - \Omega_d T \right)
\end{equation}
where $\Omega_d$ is the drift rate. The baroclinic wave appears to be stationary because it is observed from the perspective of its rotating frame, i.e., looking at the flow from the perspective of an observer that is moving at a characteristic flow speed.
This leads to an alternative way of calculating the first-order statistical moments:
$$
\overline{ \phi}^{d} \left(  r , \theta \right) = \frac{1}{T} \int_0^T \phi^{d} \left( t, r , \theta  \right) \mathrm{d}t
\label{mean}$$

and second-order statistical moments from the related fluctuations:
$$
u^{\prime d}\left(  t , r , \theta \right) = u^{\prime d}\left(  t , r , \theta \right)  -  \overline{u}^{d}\left(  r , \theta \right)
\label{fluct}$$
Figure~\ref{fig:StatisticsDriftFrame} illustrates this averaging procedure by showing a snapshot of the surface temperature, the mean, and root mean square (rms) temperature according to these definitions.

The figure on the left is an instantaneous field from our direct numerical simulation.  The middle and left figures represent the mean field according to \ref{mean} and \ref{fluct}.

\section{Harmonic analysis}\label{Appendix:HarmonicAnalysis}
Harmonic analysis is a common post-processing technique for estimating the {\it drift rate} and the {\it average temperature variance}  of baroclinic waves.  
It is based on the Fourier decomposition of the temperature in the azimuthal direction at a fixed $r,z$. Dropping the variables $r,z$ for clarity, the harmonic analysis is based on the series expansion reads:
\begin{equation*}
  T(t,\varphi) = \left< T(t, \varphi ) \right>_{\varphi} + \sum^{N_\varphi}_{p=1}A_m(t) \sin( m\varphi + \phi_m(t) )  
\end{equation*}
where $\left< \cdot \right>_{\varphi}$ is the averaged temperature over the azimuthal direction, and $m$ is the mode.
In the context of baroclinic instability, this analysis allows us to assess the flow regime according to the shape of the amplitude mode $A_m(t)$.
The drift rate is determined from the slope of the phase velocity divided by its respective mode:
$$
c_m(t)= \frac{1}{m} \frac{\mathrm{d}}{\mathrm{d}t} \phi_m(t)
$$
The {\it average temperature variance} is obtained from the time average of azimuthal standard deviation:
$$
\sigma^2 = \left< \sum^{N_\varphi}_{m=1}A^2_m(t) \right>_{t}
$$

\section{Rescaling the O'Niel marginal stability criterion}\label{Appendix:oNielStability}

\citet{oNiel1969stability} investigated the stability criteria of steady flows in rotating systems characterised by a solid bottom and free upper surfaces.
The neutral stability curve results from the solution of the transcendental equation:
\begin{equation}\label{eq:Oniel_01}
L_m(Q,\Gamma) = -\left(\Gamma^{-5/2}Ta^{-1/2}\right)^{3/2}\left(\frac{\Delta_HT}{\Delta_VT}\right)^{2}. 
\end{equation}
Here, the variable $\epsilon_b=\frac{\Delta_HT}{\Delta_VT}$ represents the temperature difference ratio between horizontal and vertical layers. In this formulation, the marginal instability curves vary with different $\epsilon_b$ values, therefore, the criteria for baroclinic wave occurrence are dependent on the vertical and horizontal temperature difference $\Delta T_V$ and $\Delta T_H$. 
The issue of assessing these two parameters is then raised. 
Examples of time-averaged temperature profiles in the $rz$ plane presented in Figure~\ref{fig:rms_temperture} show that the temperature gradient is rather constant in the core. 
Thus, a spatially averaged temperature gradient can be introduced as follows:
\begin{equation}
  \left< \nabla T \right> = \frac{1}{S} \int \nabla T \mathrm{d}S.
\end{equation}
It is possible to express $\left< \nabla T \right> $ this quantity in terms of the reference parameters of the baroclinic cavity as:
\begin{equation}
  \Delta_VT=\sigma_z \Delta T = H \left< \partial_z T \right>, \quad \Delta_HT=\sigma_r \Delta T = (b-a)\left< \partial_r T \right>
\end{equation}
We can now re-write the right-hand side of Equation~(\ref{eq:Oniel_01}) can be redefined as:
\begin{equation} \label{eq:Oniel_02}
\left(\Gamma^{-5/2}Ta^{-1/2}\right)^{3/2}\left(\frac{\Delta_HT}{\Delta_VT}\right)^{2} = \left(\epsilon_B^{8/3}Ta\right)^{-3/4}\Gamma^{-15/4} 
\end{equation}
This reformulation allows us to introduce the scaled Taylor number in equation~(\ref{eq:Oniel_01}), instead of keeping the explicit dependency on the temperature gradients. Finally, we can substitute the parameter $\epsilon_b=\Delta_VT/\Delta_HT$ by the parameterization $\epsilon_b=\sigma_z/\sigma_r$, with values presented in Figure~\ref{fig:sigma_vs_omega}.

This re-scaling presented leads to a marginal-stability curve that does not have an explicit dependency on each temperature gradient independently, i.e., it does not depend separately on $\Delta_HT$ and $\Delta_VT$, but on their ratio $\epsilon$. The curve obtained from the re-scaled Equation~(\ref{eq:Oniel_02}) is presented in Figure~\ref{fig:oneil_curve}.

\bibliography{refs}

\end{document}